\pdfoutput=1
\documentclass[twocolumn,english,showpacs,footinbib,bibnotes]{revtex4-1}

\usepackage{bm}% bold math
\usepackage{amsfonts}
\usepackage[pdftex]{graphicx}
\usepackage[usenames,dvipsnames]{color}
\usepackage{amssymb}
\usepackage{amsmath}
\usepackage{bbold}

\usepackage{bbm}
\usepackage{amsthm}
\usepackage{hyperref}
\hypersetup{
        colorlinks=false,
}
\usepackage{times}

\usepackage[english]{babel}
\usepackage{blindtext}
\usepackage{subfigure}

\newcommand{\vareps}{\varepsilon}

\newcommand{\tr}{\text{Tr}\,}
\newcommand{\subtr}{\text{Tr}}

\newcommand{\abs}[1]{\left| #1 \right|} % for absolute value

\newcommand{\ket}[1]{\left| #1 \right>}
\newcommand{\bra}[1]{\left< #1 \right|}
\newcommand{\braket}[2]{\left< #1 | #2 \right>}
\newcommand{\ketbra}[2]{\left\lvert{#2}\middle\rangle\!\middle\langle{#2}\right\rvert}
\newcommand{\mc}[1]{\ensuremath{\mathcal{#1}}}

\newcommand{\Renyi}{R\'{e}nyi\ }
\newcommand{\fcite}[1]{\ensuremath{^{[{\color{red}x}]}}}
\let\oldmarginpar\marginpar
\renewcommand\marginpar[1]{\-\oldmarginpar[\raggedleft\tiny\color{red} #1]%
{\raggedright\tiny #1}}

\newcommand{\mat}[1]{\textsf{\textbf{#1}}}
\newcommand{\mua}{\ensuremath{\mu_a}}
\graphicspath{{Figures/}}

\begin{document}

\title{Numerical stabilization of entanglement computation in auxiliary field quantum Monte Carlo simulations
	 of interacting many-fermion systems
	}
\date{\today}

\author{Peter Broecker}
\author{Simon Trebst}
\affiliation{Institute for Theoretical Physics, University of Cologne, 50937 Cologne, Germany}

%%%%%%%%%%%%%%%%%%%%%%%%%%%%%%%%%%%%%%%%%%%%%%%%%%%%%%%%%%%%%%%%%%%%%%%%%%%%%%%%%%%%%

\begin{abstract}
In the absence of a fermion sign problem, auxiliary field (or determinantal) quantum Monte Carlo (DQMC) approaches have long been the numerical method of choice for unbiased, large-scale simulations of interacting many-fermion systems. More recently, the conceptual scope of this approach has been expanded by introducing ingenious schemes to compute entanglement entropies within its framework. On a practical level, these approaches however suffer from a variety of numerical instabilities that have largely impeded their applicability. Here we report on a number of algorithmic advances to overcome many of these numerical instabilities and significantly improve the calculation of entanglement measures in the zero-temperature projective DQMC approach, ultimately allowing to reach similar system sizes as for the computation of conventional observables. We demonstrate the applicability of this improved DQMC approach by providing an entanglement perspective on the quantum phase transition from a magnetically ordered Mott insulator to a band insulator in the bilayer square lattice Hubbard model at half filling.
\end{abstract}

\maketitle

%%%%%%%%%%%%%%%%%%%%%%%%%%%%%%%%%%%%%%%%%%%%%%%%%%%%%%%%%%%%%%%%%%%%%%
% Introduction
%%%%%%%%%%%%%%%%%%%%%%%%%%%%%%%%%%%%%%%%%%%%%%%%%%%%%%%%%%%%%%%%%%%%%%

\section{Introduction}

In statistical physics, quantum Monte Carlo (QMC) simulations are a mainstay of the numerical characterization of quantum many-body systems \cite{Gubernatis2016}.
Revered for their ability to provide unbiased, numerically exact results even in the strong coupling regime, they provide valuable insight and theoretical guidance where analytical approaches such as mean-field or perturbative calculations often fail. World-line approaches such as the stochastic series expansion (SSE) \cite{SSE} (used in combination with non-local update schemes \cite{DirectedLoops,GeneralizedDirectedLoops,NonLocalUpdates}) or the continuous-time formulation of the worm algorithm \cite{Worm} have basically solved unfrustrated bosonic systems from a numerical perspective. For fermionic many-body systems, the situation is somewhat more delicate with the infamous fermion sign problem \cite{Hirsch1982,Loh1990} thwarting the polynomial efficiency of the approach for a range of problems, i.e. its power-law scaling 
with system size and inverse temperature. Nevertheless, one should not overlook that there are a number of fermionic quantum many-body systems that do {\it not} suffer from the fermion sign problem and at the same time exhibit interesting physics arising solely from the fermionic nature of its constituents. This includes, for instance, the formation of superconductivity in microscopic models of nearly antiferromagnetic metals \cite{Berg2012,Schattner2016}, the emergence of topological order in Kane-Mele-Hubbard models \cite{Bercx14,Hohenadler14,Holdin15}, or fermionic quantum critical phenomena in Dirac matter  
\cite{Wang2014,Li2015,Motruk2015,Capponi2015,Broecker2016} or Z$_2$ gauge field coupled fermion systems \cite{AssaadGrover2016,Vishwanath2016}. 
Large-scale lattice models of all these fermionic systems have been simulated using DQMC simulations in the past three years, typically
with an emphasis on the statistical measurement of observables such as order parameters, correlation functions, or topological invariants.
At the same time, conceptual advancements have paved the way to calculate entanglement measures such as \Renyi entropies or entanglement spectra within the framework of DQMC simulations \cite{Grover2013,Broecker2014,Assaad2014,Assaad2015}, in step with similar developments for their bosonic world-line counterparts \cite{Hastings2010,Herdman2014}.
However, the practical use of these entanglement computations has so far been impeded by a number of numerical instabilities arising deep
within the DQMC technique. The purpose of the manuscript at hand is to give a detailed account as to how to overcome these numerical insufficiencies and bring the computation of entanglement measures en par with conventional observables.
Our algorithmic improvements focus on the calculation of \Renyi entropies using an adaptation of the replica scheme \cite{holzhey_geometric_1994,Calabrese2004}
to the DQMC approach \cite{Broecker2014}. The numerical stabilization of its computation includes a variety of algorithmic steps ranging from linear algebra aspects such as the inversion of seemingly singular matrices and technical tricks such the inclusion of an artificial chemical potential, which is shown to control the condition numbers of the underlying linear algebra algorithms, to more DQMC specific aspects such as the optimal choice of Hubbard-Stratonovich transformation and respective auxiliary fields. Beyond the numerical stabilization we provide a detailed discussion of the convergence behavior of our enhanced algorithm. 
Finally, to illustrate the performance of the technical improvements laid out in this manuscript, we apply our  DQMC + replica scheme framework to provide an entanglement perspective on the quantum phase transition  between an antiferromagnetically ordered Mott insulator and a featureless band insulator in the bilayer Hubbard model on the square lattice.

The remainder of the manuscript is organized as follows. We start with a brief review of the general auxiliary field quantum Monte Carlo approach in Sec.~\ref{sec:DQMC} where we put a particular emphasis on those aspects where numerical instabilities typically arise and how they are dealt 
with in the conventional algorithm \cite{blankenbecler_monte_1981,santos_introduction_2003-1,assaad_worldline_2008} that is used to compute correlation functions at finite temperatures.
We then turn to the replica scheme and discuss its conceptual adaptation to the DQMC framework in Sec.~\ref{sec:Replica}. The main results of the manuscript are provided in Sec.~\ref{sec:Stabilization} where we discuss the numerical stabilization for the calculation of \Renyi entropies, followed by a discussion of the optimal choice of Hubbard-Stratonovich transformation in Sec.~\ref{sec:HS} and the overall convergence behavior of the algorithm in Sec.~\ref{sec:Convergence}. The application of the so-improved DQMC approach to the quantum phase transition in the bilayer Hubbard model is presented in Sec.~\ref{sec:Bilayer}. We close with a discussion in Sec.~\ref{sec:Conclusions}.

%%%%%%%%%%%%%%%%%%%%%%%%%%%%%%%%%%%%%%%%%%%%%%%%%%%%%%%%%%%%%%%%%%%%%%
% Determinantal QMC
%%%%%%%%%%%%%%%%%%%%%%%%%%%%%%%%%%%%%%%%%%%%%%%%%%%%%%%%%%%%%%%%%%%%%%

\section{Auxiliary field quantum Monte Carlo}
\label{sec:DQMC}

We start our discussion with a brief recapitulation of the basics of the DQMC method \cite{blankenbecler_monte_1981}. We will primarily
discuss its finite-temperature formulation and only touch on some of the aspects of the alternative zero-temperature projective scheme. 
This choice of focus is motivated by the fact that the algorithm for the computation of ground-state entanglement properties is conceptually
much closer to the finite-temperature approach than the projective scheme. Readers looking for a more exhaustive introduction to DQMC
approaches are referred to the review articles in Refs.~\onlinecite{santos_introduction_2003-1, assaad_worldline_2008}.

%%%%%%%%%%%%%%%%%%%%%%%%%%%%%%%%%%%%%%%%%%%%%%%%%%%%%%%%%%%%%%%%%%%%%%
% Conceptual formalism
%%%%%%%%%%%%%%%%%%%%%%%%%%%%%%%%%%%%%%%%%%%%%%%%%%%%%%%%%%%%%%%%%%%%%%
\subsection{Conceptual formalism}
\label{sec:Conceptual}

From a conceptual perspective, the goal of determinantal quantum Monte Carlo approaches is to sample the partition function 
\begin{equation}
	Z = \tr {e^{-\beta \hat{H}}} \,.
	\label{eq:Z}
\end{equation}
for a given lattice Hamiltonian $\hat{H}$ in the grand-canonical ensemble and to evaluate  finite-temperature expectation values of correlation functions along the way.
The first step in accomplishing this is to perform a Trotter-Suzuki decomposition, discretizing the inverse temperature or imaginary time into $N_\tau$ time slices of size $\Delta\tau = \beta / N_\tau$. 
The investigated Hamiltonians are typically of the form $\hat{H} = \hat{K} + \hat{V}$ with a non-interacting, quadratic part $\hat{K}$  and an interacting, quartic part $\hat{V}$, respectively.
By making use of the Baker-Campbell-Hausdorff formula, one can split the sum in the exponential into a product of exponentials and keep only the lowest order term
\begin{equation}
Z = \tr \prod\limits_{\tau=1}^{N_\tau} e^{-\Delta\tau \hat{K}} e^{-\Delta\tau \hat{V}} + \mc{O}(\Delta\tau^2) \,.
\end{equation}
Ignoring the higher orders introduces an error of size $\mc{O}(\Delta\tau^2)$ that can be considered negligible if $\Delta\tau$ is sufficiently small.
To evaluate the interaction part, we apply a Hubbard-Stratonovich transformation~\cite{stratonovich_method_1957, hubbard_calculation_1959} on each time slice and each term in the interaction operator reducing quartic to quadratic operators at the cost of introducing an auxiliary field that we denote by $s(\tau, j)$, with $\tau$ being the time slice and $j$ the index indicating the decoupled term of the interaction operator.
After this Hubbard-Stratonovich transformation, we end up with a system of {\it free} fermions coupled to the auxiliary field via an interaction $V(s, \tau)$  that depends on the auxiliary field configuration and the time slice in imaginary time.
The operators $\hat{K}$ and $\hat{V}(s, \tau)$ are now simply one-particle operators and can be written out in their matrix form $\mat{K}$ and $\mat{V}(s, \tau)$, respectively, so that the partition sum is given as
\begin{align}
 Z &= \sum\limits_{s(\tau, j)} \tr \prod\limits_{\tau=1}^{N_\tau} e^{-\Delta\tau \mat{K}} e^{-\Delta\tau \mat{V}(s, \tau)} \nonumber  \\
 &= \sum\limits_{s(\tau, j)} \tr \prod\limits_{\tau=1}^{N_\tau}\mat{B}(s, \tau) \nonumber \\
\end{align}
where $\mat{B}(s, \tau)$ denotes the matrix product of the hopping matrix and the decoupled interaction matrix for a given time slice $\tau$ and a given auxiliary field configuration $s$.
Let us introduce some notation that we will use throughout the following. 
First, in favor of a clearer notation, we will the drop the auxiliary field argument $s$, since we will focus on calculating objects for a fixed configuration. 
We define a partial product of slice matrices starting at time slice $\tau$ and ending at time slice $\tau^\prime$ as
\begin{equation}
	B(\tau^\prime, \tau) = \prod\limits_{i=\tau}^{\tau^\prime} \mat{B}(i) \,,
	\label{eq:PartialMatrixProduct}
\end{equation} 
The full product over all time slices is denoted as
\begin{equation}
	\mc{B}(\tau = 1) = \prod\limits_{i=1}^{N_\tau} \mat{B}(i) \,,
	\label{eq:MatrixProduct}
\end{equation} 
where the label $\tau$ in the latter indicates the starting point for the matrix product.
If $\tau \neq 1$, then the corresponding matrix product should be viewed as cyclically permuted by $\tau - 1$ elements,
a perspective which we will later use when accessing equal-time Green's functions at different imaginary time slices.
Using this notation, we may write the partition sum in the following concise way
\begin{equation}
Z = \sum\limits_{\{s(\tau, j)\}} \tr \, \mc{B}(\tau = 1) \,.
\end{equation}
At this point, it should be noted that the partition sum still contains a trace over all fermion states \emph{and} over all auxiliary field configurations $s(\tau, j)$.
We can, however, integrate out the free fermions, resulting in one {\it determinant} per auxiliary field and fermion configuration.
It can further be shown \cite{santos_introduction_2003-1, assaad_worldline_2008} that the sum of determinants reduces to a {\it single} determinant quantifying the electron contribution to the partition sum for any given configuration of the auxiliary field. This allows to reformulate the original partition sum \eqref{eq:Z} in terms of determinants as 
\begin{equation}
Z = \sum\limits_{\{s(\tau, j)\}} \det{(1 + \mc{B}(\tau))} \,.
\end{equation}
A central object of quantum statistical physics is the equal-time Green's function $G(\tau)$, which is easily accessible in the DQMC algorithm
with its matrix form related to the above determinants via
\begin{equation}
G(\tau) = (1 + \mc{B}(\tau))^{-1}\label{eq:greens_b} \,.
\end{equation}
The equal-time Green's function not only provides valuable access to correlation functions, it also plays a crucial role in devising an efficient updating procedure of the auxiliary field configurations sampled at the heart of the Monte Carlo approach. For details we again refer to the  literature \cite{santos_introduction_2003-1, assaad_worldline_2008}.

%%%%%%%%%%%%%%%%%%%%%%%%%%%%%%%%%%%%%%%%%%%%%%%%%%%%%%%%%%%%%%%%%%%%%%
% Numerical implementation
%%%%%%%%%%%%%%%%%%%%%%%%%%%%%%%%%%%%%%%%%%%%%%%%%%%%%%%%%%%%%%%%%%%%%%

\subsection{Numerical implementation}
\label{sec:Numerics}

Any practical implementation of the algorithm described above typically faces two problems, both related to the underlying linear algebra:
(i) the evaluation of the matrix product \eqref{eq:MatrixProduct} is found to be numerically unstable~\cite{Bai2011659} and (ii) the resulting matrix $\mc{B}(\tau)$ turns out to be ill-conditioned, which hampers both the calculation of the Monte Carlo transition probabilities as well as the evaluation of the equal-time Green's function and any observables derived from it. 
The condition number of the matrix product increases with increasing hopping strength and increasing inverse temperature, respectively and may increase or decrease with increasing interaction strength depending on the type of Hubbard-Stratonovich transformation.

Stabilizing a matrix product of the form \eqref{eq:MatrixProduct} is a well understood procedure in numerical mathematics.
The idea is to perform either a rank revealing QR or a singular value decomposition (SVD) after a certain number of multiplications $m$, typically chosen small enough such that this shorter product remains exact up to machine precision. 
By making use of one of the decompositions one recasts a given matrix $M$ into a product $M = U D T$, where $U$ is typically unitary, $D$ is a diagonal matrix that contains the information about the inherent scales of the matrix (e.g. in the form of its singular value spectrum), and $T$ is either unitary (SVD) or triangular (QR decomposition). Independent of the specific decomposition algorithm, we will refer to the values of the diagonal matrix $D$ as the singular values.
To stabilize the computation of the matrix product \eqref{eq:MatrixProduct},  we divide the imaginary time interval into $N_m$ groups of $m$ time slices representing a segment of length $\Delta = m\cdot \tau$ in imaginary time that encompasses $m$ slice matrices $\mat{B}(i)$.
After the multiplication and decomposition of the first $m$ matrices into $U_1 D_1 T_1$,  the next set of matrices is then multiplied only to $U_1$, {\em before} $D_1$ is multiplied from the right and the resulting product is decomposed again, i.e. 
\begin{equation}
\left(\left(\prod\limits_{i = m + 1}^{2m} \mat{B}(i)\cdot U_1\right)D_1\right)T_1 = U_2 D_2 (T T_1) = U_2 D_2 T_2 \,.
\end{equation}
This procedure is repeated until all slice matrices are incorporated into the $U D T$ decomposition. 
Note that we will use all the incomplete, intermediate results at a later point so they will be stored in memory. 
We refer to this set of matrices, $\{\{U_i\}, \{D_i\}, \{T_i\}\}$,  as the {\it stack}.

The determinant that gives the weight of a specific configuration of the auxiliary field is given as
\begin{equation}
	\det{\left(\mathbb{1} + \mc{B}(\tau) \right)} = \det{\left(\mathbb{1} + UDT\right)}\label{eq:det_weight} \,,
\end{equation}
and the equal-time Green's function as
\begin{equation}
	G(\tau) = \left(\mathbb{1} + \mc{B}(\tau)\right)^{-1} = \left(\mathbb{1} + UDT\right)^{-1} \,.
	\label{eq:greens}
\end{equation}
The addition of the identity matrix, although seemingly simple, requires careful attention to ensure an accurate calculation of the determinant and the Green's function.
This problem is solved by separating the diagonal matrix $D$ from the auxiliary matrices $U$ and $T$ before the resulting matrix is decomposed yet again:
\begin{equation}
\left(\mathbb{1} + UDT\right) = U\left(U^{-1} T^{-1} + D\right) T =  \left(U U^\prime\right) D^\prime  \left( T^\prime T \right) \,.
\label{eq:weight}
\end{equation}
The weight, i.e. the determinant, can then easily be read off as the product of the diagonal entries of $D^\prime$. 
It should be noted that in practice one would calculate the logarithm of the determinant to avoid numerical overflow and rounding errors.
\begin{figure}
\includegraphics[width=\columnwidth]{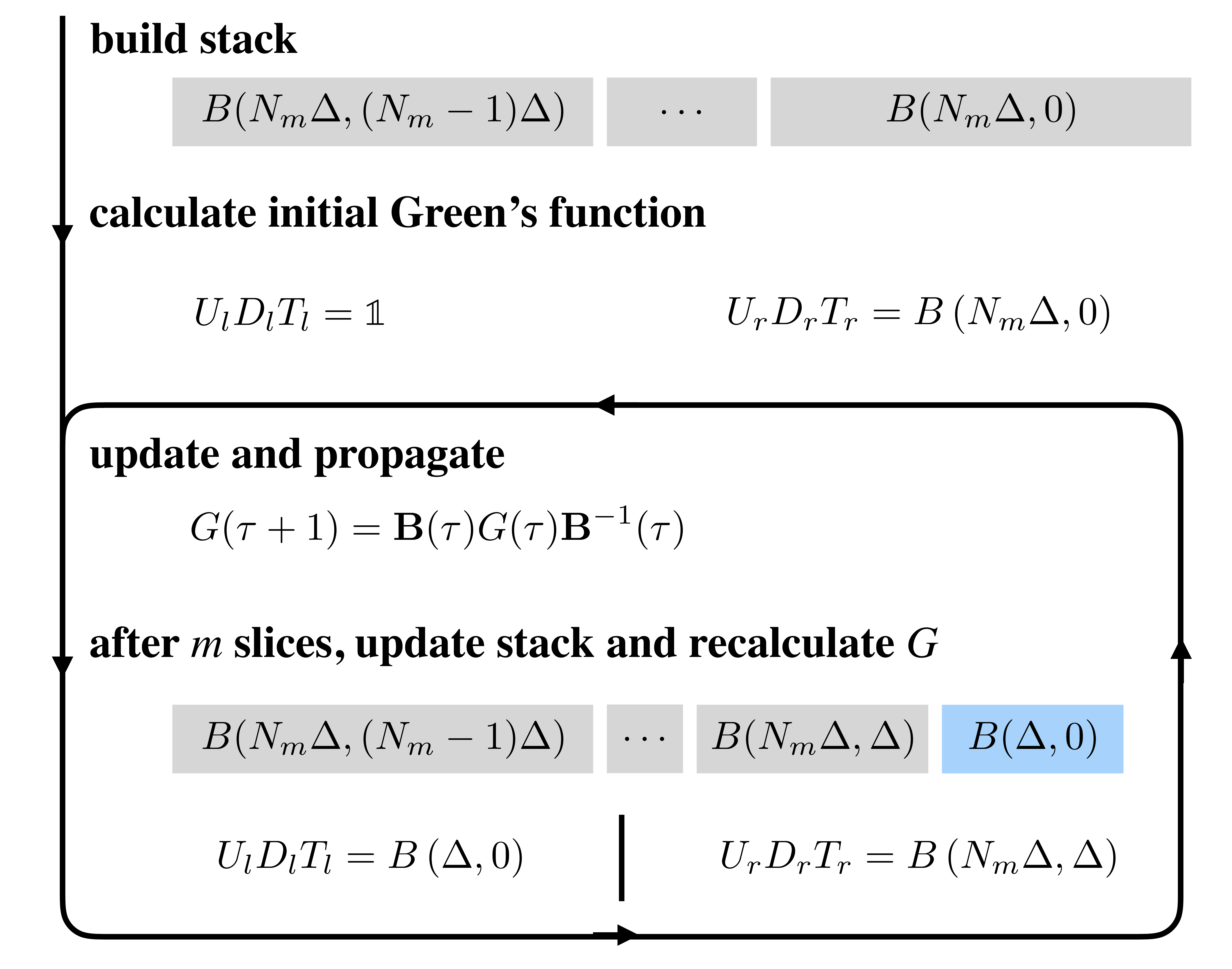}
\caption{Flow diagram of the basic DQMC algorithm. Shaded in grey is the initial build up of the stack of decomposed matrix products. This is used to calculate the Green's function which is subsequently updated and propagated in imaginary time before updating the stack and recalculating the Green's function from the stack to ensure stability. The slice product shaded in red has to be recalculated from scratch because it contains auxiliary field values that were updated in the Monte Carlo procedure. Subsequent slice products will the respective preceding $U$ matrix as a starting point for the recalculation.
\label{fig:dqmc_scheme}
}
\end{figure}

We already mentioned that the Green's function at a given imaginary time $\tau$ is also used in the updating procedure of the auxiliary field for the corresponding time slice, which means that we need to have access to the Green's function at each time slice. 
One option would be to recalculate the Green's function at every time slice from scratch, but that is numerically very expensive. Instead, one may propagate the Green's function along imaginary time by virtue of
\begin{equation}
G(\tau + 1) = \mat{B}(\tau) G(\tau ) \mat{B}^{-1}(\tau)\label{eq:greens_propagation} \,.
\end{equation}
This equation suggests that we have to calculate the Green's function only once, e.g. at $\tau = 1$, and then continue propagating the Green's function along imaginary time updating the appropriate auxiliary field degrees of freedom. 
However, the propagation suffers from the same problem as the build up of the matrix product \eqref{eq:MatrixProduct} itself and is stable only for a few steps.
In practice, we assume the number of stable propagation steps $N_{\text{prop}}$ to be equal to the number of slices $m$ making up groups of $\mat{B}$-matrices.
In general, this number can be determined by considering the difference between the propagated Green's function and the recalculated one after $m$ propagations have taken place. The element-wise relative deviation 
\begin{equation}
\delta_{ij} = 2\cdot \dfrac{\abs{G^{\text{recalculated}}_{ij} - G^{\text{propagated}}_{ij}}}{\abs{G^{\text{recalculated}}_{ij} + G^{\text{propagated}}_{ij}}}\label{eq:greens_error}
\end{equation}
between the two should be small enough, such that the update probabilities and measurements are not seriously affected. Therefore, one typically focuses on comparing only those matrix elements that are used in the updating and measurement procedure. 
For those imaginary time slices where the Green's function has to be recalculated from scratch, one can greatly facilitate its calculation by making use of the matrix products saved in the stack. 

We summarize the basic DQMC algorithm in the flow diagram provided in Fig.~\ref{fig:dqmc_scheme}.
%

%%%%%%%%%%%%%%%%%%%%%%%%%%%%%%%%%%%%%%%%%%%%%%%%%%%%%%%%%%%%%%%%%%%%%%
% The replica trick
%%%%%%%%%%%%%%%%%%%%%%%%%%%%%%%%%%%%%%%%%%%%%%%%%%%%%%%%%%%%%%%%%%%%%%

\section{The replica scheme}
\label{sec:Replica}

To characterize and quantify the entanglement in a many-fermion system, we consider entanglement entropies \cite{Bekenstein1973,Hawking1975} calculated for a bipartition of the system into two complementary subsystems $A$ and $B$. Entanglement entropies have become a standard measure of entanglement for quantum many-body systems that can reveal valuable information about the principal nature of a quantum state \cite{Laflorencie2016}. 
Such a classification of quantum states is typically achieved by a careful scaling analysis of the entanglement entropies with system size, which reveals a dominant scaling with the length of the boundary of the bipartition \cite{Eisert2010} and several subleading contributions revealing e.g. the formation of macroscopic, long-range entanglement in topologically ordered systems \cite{LevinWen,KitaevPreskill}, the formation of more conventionally symmetry-broken ordered states via a detection of the corresponding Goldstone modes \cite{MetlitskiGrover2011,Kulchytskyy2015} or the existence of a Fermi surface via a logarithmic violation of the boundary law \cite{Wolf2006,Klich2006}.
Technically, entanglement entropies are typically calculated in the form of \Renyi entropies, which can be computed via the so-called replica scheme, a well-known procedure originally introduced in the context of quantum field theories~\cite{holzhey_geometric_1994}. 
In the following, we will quickly review how the replica scheme can be adapted to the framework of DQMC techniques.

%%%%%%%%%%%%%%%%%%%%%%%%%%%%%%%%%%%%%%%%%%%%%%%%%%%%%%%%%%%%%%%%%%%%%%
% Renyi entropies
%%%%%%%%%%%%%%%%%%%%%%%%%%%%%%%%%%%%%%%%%%%%%%%%%%%%%%%%%%%%%%%%%%%%%%

\subsection{\Renyi entropies}
\label{sec:Renyi}

\Renyi entropies constitute a family of entanglement entropies calculated from the reduced density matrix $\rho_A = \subtr_B \rho$ for a bipartition of the system into two complementary subsystems $A$ and $B$
\begin{equation}
  S_n = \dfrac{1}{1 - n}   \log{ \tr{ \left( \rho_A^n \right) }} \,,
  \label{eq:renyi}
\end{equation}
where the index $n$ is typically an integer number.
This family of \Renyi entropies includes the well-known von Neumann entropy $S = -\tr \left( \rho_A \log \rho_A \right)$, which is recovered
in the limit of $n\rightarrow 1$. For the purpose of our numerical Monte Carlo simulations we will, however, consider only \Renyi entropies
with $n > 1$, typically considering the case of the second \Renyi entropy $n=2$, for which one can formulate the replica scheme~\cite{holzhey_geometric_1994}.

\begin{figure}[t]
\includegraphics[width=\columnwidth]{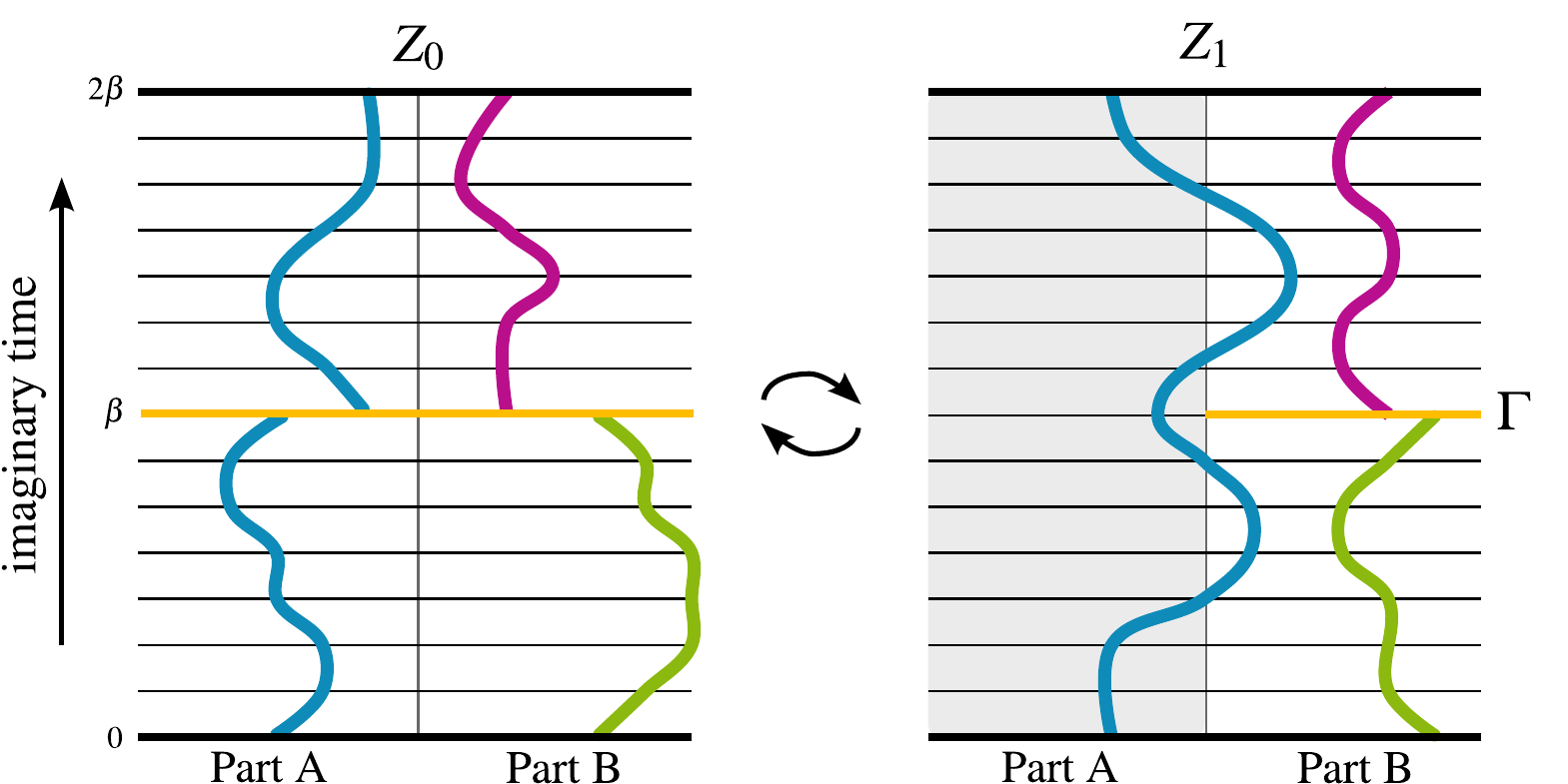}
\caption{(Color online) Worldline representation of the two partition sums of \eqref{eq:renyi_two}. On the left side, the squared, regular partition function is depicted. Worldlines of particles are $\beta$-periodic, no matter what subsystem they originate from. The right hand side shows the modified partition sum that is $\beta$-periodic in part $B$ but $2\beta$-periodic in part $A$. \label{fig:ensemble_switching_paths}}
\end{figure}

To quickly motivate and rederive this scheme let us start with the density matrix $\rho$ of the quantum many-body system at hand. In general,
we can think of this density matrix as being subject to a normalization via a factor $\tr{\rho}$, which is, of course, unaffected if one takes the partial trace over degrees of freedom of part $B$. Inserting this into Eq.~\eqref{eq:renyi} gives
\begin{equation}
  S_n = \dfrac{1}{1 - n}   \log{ \dfrac{\tr{ \left( \rho_A^n \right) }}{\left( \tr\rho\right)^n}} \,,
  \label{eq:renyi_reduced}
\end{equation}
which we rewrite as the {\it ratio of two partition sums}
\begin{equation}
  S_n = \dfrac{1}{1 - n} \log{\dfrac{\mc{Z}[A, n, \beta]}{\mc{Z}^n}} \,,
\end{equation}
with $\mc{Z}[A, n, \beta] = \tr{ \left( \rho_A^n \right) } $ and $\mc{Z}^n = \left( \tr\rho\right)^n$, respectively. 
In the following, let us focus on the second \Renyi entropy, i.e. $n=2$, for which the two equations above reduce to
\begin{equation}
  S_2 = -  \log{ \dfrac{\tr{ \left( \rho_A^2 \right) }}{\left( \tr\rho\right)^2}} = -\log{\dfrac{\mc{Z}[A, 2, \beta]}{\mc{Z}^2} \equiv -\log{\dfrac{Z_1}{Z_0}}}  \,.
  \label{eq:renyi_two}
\end{equation}
Note that $\mc{Z}$ is the usual partition function, which we sample in conventional DQMC simulations to obtain numerical estimates of conventional observables such as, for instance, correlation functions. 
The second partition function $\mc{Z}[A, 2, \beta]$ entering the above equation is a modified partition function, where part $B$ was traced out and the resulting reduced density matrix was squared ($n = 2$) before the final trace over $A$ was taken.
Associated to these partition functions are Green's functions $G_0$ and $G_1$, respectively.
Written out explicitly $\mc{Z}[A, 2, \beta]$ takes the form
\begin{align}\label{eq:z_num}
  \tr_A \left(\rho_A^{\prime 2} \right) &= \sum\limits_{\mathcal{A}, \mathcal{A}^\prime, \mathcal{B}, \mathcal{B}^\prime}
  \bra{\mathcal{A}{\mathcal{B}^\prime}}
  \rho^\prime
  \ket{\mathcal{A}^\prime{\mathcal{B}^\prime}}
  \bra{\mathcal{A}^\prime{\mathcal{B}}}
  \rho^\prime
  \ket{\mathcal{A}\mathcal{B}}\nonumber \\
  &\equiv \mc{Z}[A, 2, \beta] \,.
\end{align}
Note that $\mc{Z}[A, 2, \beta]$  is $\beta$-periodic in part $B$ which due to the squaring appears twice, once as $\mc{B}$ and once as $\mc{B}^\prime$, while part $A$ is $2\beta$-periodic and appears only once.
These peculiar boundary conditions in imaginary time can best be understood when visualized in a world-line picture as shown in Fig.~\ref{fig:ensemble_switching_paths}.

%%%%%%%%%%%%%%%%%%%%%%%%%%%%%%%%%%%%%%%%%%%%%%%%%%%%%%%%%%%%%%%%%%%%%%
% The replica trick in ground state simulations
%%%%%%%%%%%%%%%%%%%%%%%%%%%%%%%%%%%%%%%%%%%%%%%%%%%%%%%%%%%%%%%%%%%%%%

\subsection{The replica scheme in DQMC simulations}

\begin{figure}[t]
\includegraphics[width=\columnwidth]{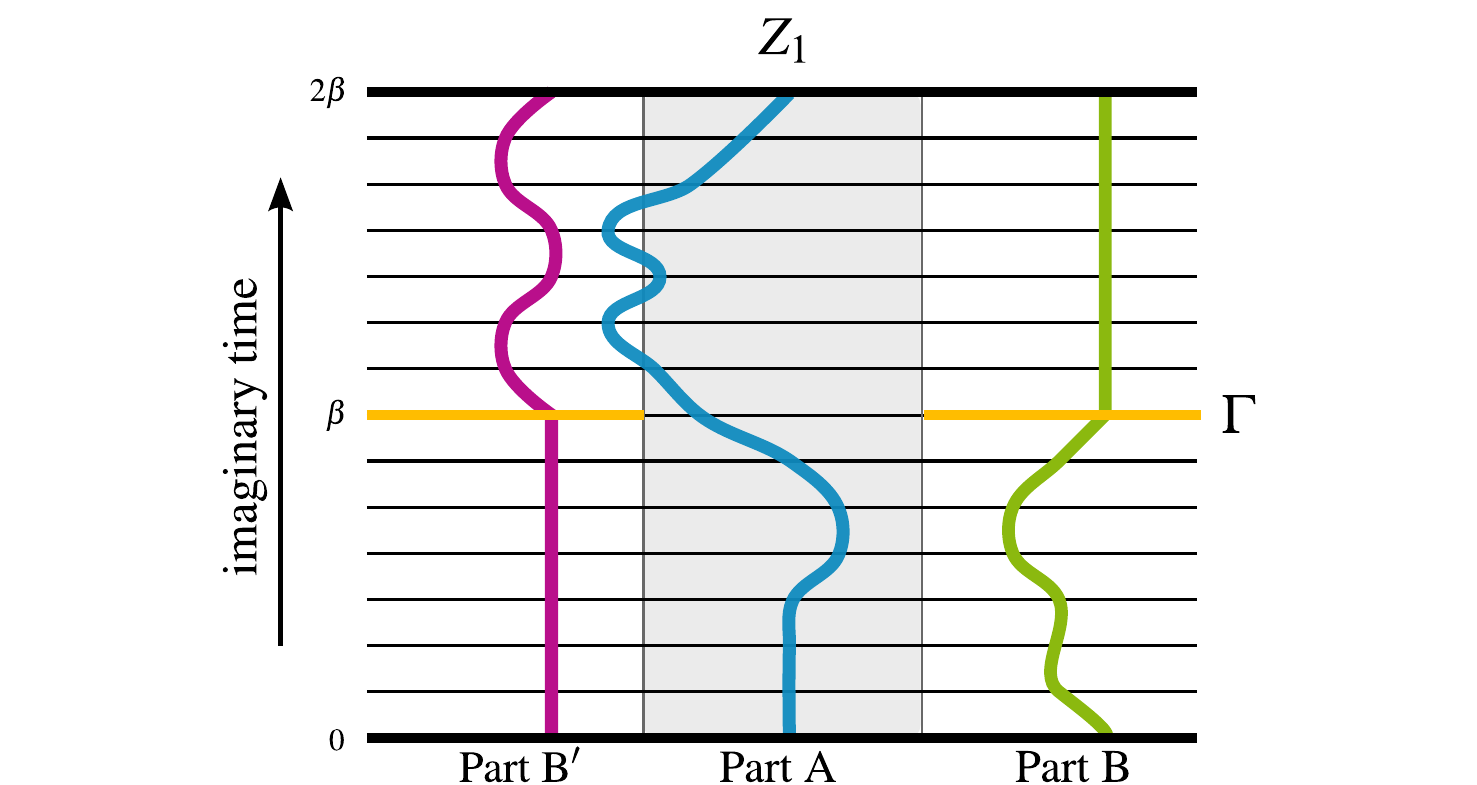}
\caption{ (Color online) Modified simulation cell used in DQMC simulations. Part $B$ appears twice as $B$ and $B^\prime$ and both extend from $0$ to $2\beta$ but its degrees of freedom only propagate in an interval of length $\beta$. \label{fig:ensemble_switching_unfolded}}
\end{figure}

To adapt this replica scheme in DQMC simulations, we have previously shown \cite{Broecker2014} that one can ``fold out" subsystems $B$ and $B'$ and consider a modified Hamiltonian that explicitly depends on imaginary time and takes the form
\begin{equation}\label{eq:img_time_ham}
\widetilde{\mathcal{H}}(\tau) = \mathcal{H}_{AB}\;\Theta(\tau)\;\Theta(\beta - \tau) + \mathcal{H}_{AB^\prime}\;\Theta(\tau-\beta)\;\Theta(2\beta - \tau) \,.
\end{equation}
Note that the simulation of this modified Hamiltonian also requires a modified simulation cell, where, according to Eq.~\eqref{eq:img_time_ham}, the degrees of freedom in subsystem $A$ interact with those in $B$ and $B^\prime$ not over the entire span of imaginary time but only in certain intervals as visualized in Fig.~\ref{fig:ensemble_switching_unfolded}.
We will refer to the subsystem $B$ or $B^\prime$ that is currently in interaction with $A$ as the {\it active subsystem}. 
In practice, the matrices associated with hopping and interaction processes are of size $(N_A + 2\cdot N_B)\times (N_A + 2\cdot N_B)$ where operators whose support lies exclusively in the inactive subsystem are represented by identity matrices and those mixing two subsystems are zero. 

The replica scheme, as introduced above, has been formulated in the context of {\it finite-temperature} simulations with the inverse temperature $\beta$ entering as a natural parameter in the replica scheme. We now want to adopt this approach to the {\it zero-temperature} projective DQMC algorithm that aims at directly sampling the ground-state wavefunction (in lieu of the partition sum). As will become clear in the following, this adaptation will result in an algorithm that is much closer to the finite-temperature scheme introduced above than the original ground-state algorithm~\cite{sorella_novel_1989}.
At the heart of the ground-state algorithm is a projective scheme applied to a trial wave function $\ket{\psi_T}$
\begin{equation}
	 \ket{\psi} = \lim\limits_{\theta \rightarrow \infty} e^{-\theta \mc{H}}\ket{\psi_T} \,,
	 \label{eq:projection_main}
\end{equation}
which for sufficiently long projection times $\Theta$ yields the desired ground-state wavefunction $\ket{\psi}$.
The only requirement imposed on the trial wave function is for it to have a non-zero overlap with the actual ground-state wavefunction, in which case the projection will eliminate all contributions from excited states and converge to the ground-state wavefunction in the limit of $\theta\rightarrow\infty$.
Again, we can introduce a normalization constant -- similar to the normalization of the density matrix in the discussion preceding Eq.~\eqref{eq:renyi_reduced} above, which can again be written as a trace over the ground state density matrix
\begin{equation}
\mc{N} = \braket{\psi}{\psi} = \sum\limits_{\ket{\mc{A}}} \braket{\psi}{\mc{A}}\braket{\mc{A}}{\psi} = \tr \left(\ket{\psi}\bra{\psi}\right) \,.
\end{equation}
We can readily insert this normalization constant (or trace) into the denominator of the definition of the \Renyi entropy in Eq.~\eqref{eq:renyi_two}.
Inserting the projection \eqref{eq:projection_main} into the nominator of the definition of the \Renyi entropy in \eqref{eq:renyi_two}, we find an expression for the $\tr \rho_A^{\prime\,2}$ that looks very similar to the finite temperature expression discussed above
\begin{widetext}
\begin{align}\label{eq:replica_gs}
\tr \,\rho_A^{\prime\,2} = \lim\limits_{\theta \rightarrow \infty} 
\sum\limits_{\mathcal{A}, \mathcal{A}^\prime, \mathcal{B}, \mathcal{B}^\prime}
\bra{\mathcal{A}{\mathcal{B}^\prime}}
\exp\left({-\theta\mc{H}}\right)\ketbra{\psi_T}{\psi_T}\exp\left({-\theta\mc{H}}\right)
\ket{\mathcal{A}^\prime{\mathcal{B}^\prime}}
\bra{\mathcal{A}^\prime{\mathcal{B}}}
\exp\left({-\theta\mc{H}}\right)\ketbra{\psi_T}{\psi_T}\exp\left({-\theta\mc{H}}\right)
\ket{\mathcal{A}\mathcal{B}} \,,
\end{align}
\end{widetext}
At first look, the only difference appears to be the occurrence of the density matrices $\ketbra{\psi_T}{\psi_T}$.
However, note that the finite temperature algorithm works in a grand-canonical ensemble, while the trace above is given in a canonical ensemble.
Nevertheless, we may safely adopt all of the machinery to implement the replica scheme described above, since the density matrices act as projectors onto the correct particle sectors.

%%%%%%%%%%%%%%%%%%%%%%%%%%%%%%%%%%%%%%%%%%%%%%%%%%%%%%%%%%%%%%%%%%%%%%
% Numerical stabilization
%%%%%%%%%%%%%%%%%%%%%%%%%%%%%%%%%%%%%%%%%%%%%%%%%%%%%%%%%%%%%%%%%%%%%%

\section{Numerical stabilization}
\label{sec:Stabilization}

With the conceptual framework for the computation of \Renyi entropies in DQMC simulations laid out in Secs. \ref{sec:DQMC} and \ref{sec:Replica} above, we now
turn to the technical aspects of its implementation. In particular, we discuss the numerical stabilization procedures enabling us to compute expectation values for these entanglement entropies for similar system sizes as for conventional observables (such as correlation functions) using the ordinary DQMC approach.

%%%%%%%%%%%%%%%%%%%%%%%%%%%%%%%%%%%%%%%%%%%%%%%%%%%%%%%%%%%%%%%%%%%%%%
% invertibility
%%%%%%%%%%%%%%%%%%%%%%%%%%%%%%%%%%%%%%%%%%%%%%%%%%%%%%%%%%%%%%%%%%%%%%

%
\begin{figure}[b]
\includegraphics[width=\columnwidth]{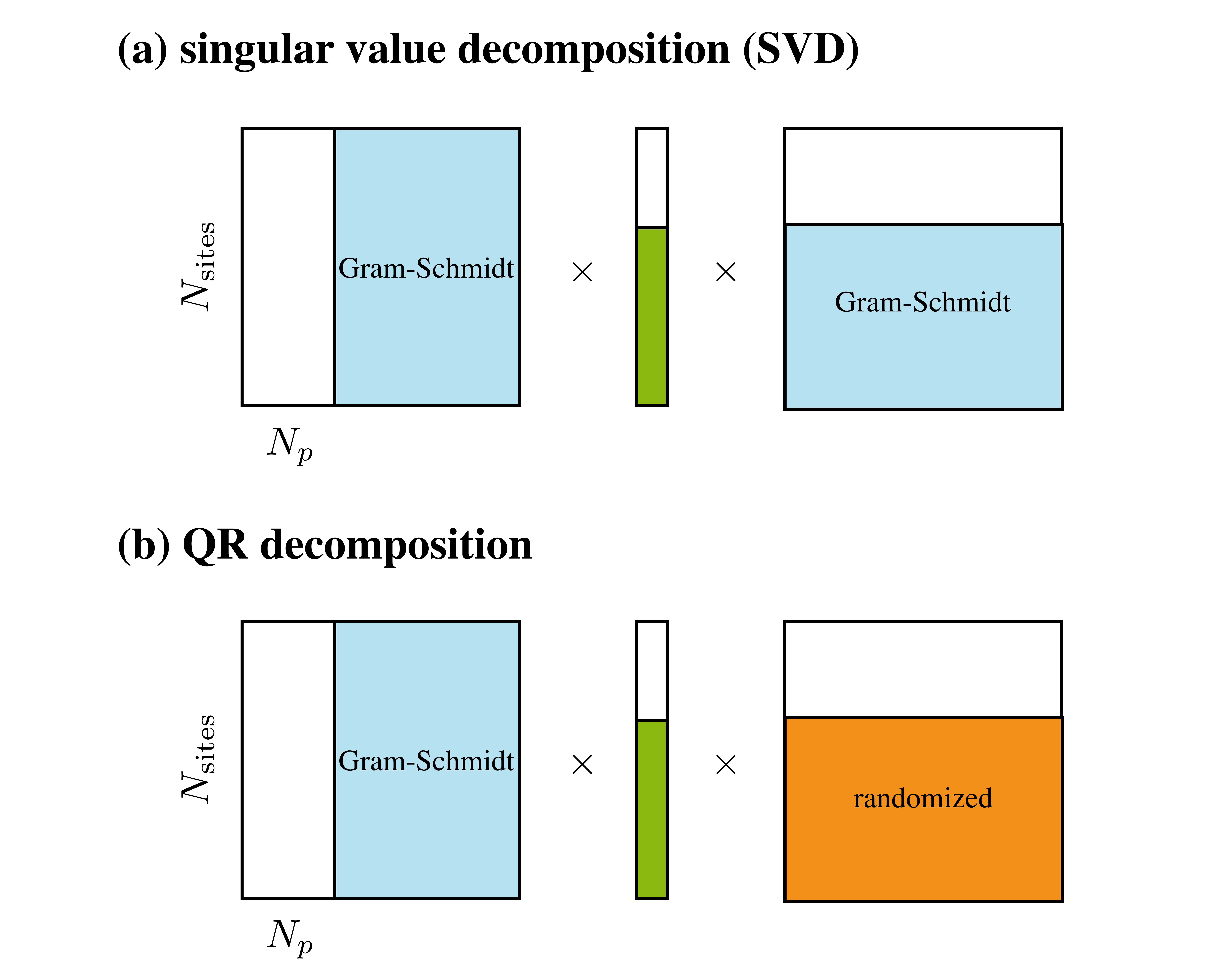}
\caption{(Color online) Inversion scheme as applicable to SVD (top) and QR (bottom) based Green's function calculations. Shown here is the example of the projected trial density matrix $B(\tau, \tau / 2)\ket{\psi_T}\bra{\psi_T}B(\tau / 2, 0)$. 
Matrices with orthogonal columns or rows that form an incomplete basis are extended to a full basis using a Gram-Schmidt process, denoted here in blue.
The singular value vectors are extended by a vector of decreasing numbers clearly separated from the smallest physical singular value by the singular value gap $\Delta$ here pictured in green. 
For the case of the QR decomposition matrices without orthogonal columns or rows appear naturally which are extended by a random matrix to match dimensions and ensure invertibility (for example by LU decomposition), pictured in orange.
\label{fig:invertibility_ground_state}}
\end{figure}

\subsection{Invertibility of the matrix products}
\label{sec:invertibility}

A fundamental problem in the adoption of the finite-temperature algorithm for ground-state properties is that the density matrix typically turns out to be  singular, i.e. it is in general not invertible. As a consequence, the entire matrix product $\mc{B}(\tau)$ and its decomposed form \eqref{eq:greens}  are also no longer invertible, which ultimately prevents to compute the equal-time Green's function and the associated Monte Carlo weights.
However, one can overcome this algebraic obstacle by numerically constructing a matrix that is equal to the matrix product $\mc{B}(\tau)$ up to machine precision, but whose matrix decomposition still yields invertible matrices as discussed in the following.

To see that the density matrix is in general a singular, non-invertible matrix let us consider a scenario where $N_p$ particles populate a lattice of $N$ sites. A trial wave function for such a scenario is typically constructed by diagonalizing the quadratic part of the Hamiltonian and then keeping only as many eigenvectors as the number of particles $N_p$ in the system \cite{Footnote:Degeneracy}.
This test wave function is thus represented as a matrix $\psi_T$ of dimension $N \times N_p$ with orthogonal columns, whereas
the associated density matrix $\rho$ of size $N\times N$ is constructed as $\rho = \psi_T \psi_T^\dagger$.
From this construction it becomes apparent that the density matrix must be {\it rank deficient} (since it is the product of two non-square matrices)
and as such singular.

To elude this problem, we modify the matrices $U, D$, and $T$ of the singular value decomposition \eqref{eq:weight} such that they become invertible but leave the original matrix unchanged (up to machine precision, which we denote by $\vareps$). 
In Fig.~\ref{fig:invertibility_ground_state}, we illustrate how to do this in practice: Any non-square matrix with orthogonal (unitary) columns (or rows) can be extended to a square, fully orthogonal (unitary) matrix by applying a Gram-Schmidt process. To ensure that the original matrix, i.e. the one that results from remultiplying $U\,D\,T$, remains unchanged, we extend the diagonal matrix by values that are at least $1/\vareps$ smaller in magnitude than the smallest singular value found in the original $D$ matrix. Numerically, this means that none of the additional columns or rows actually contribute, because they are weighted below numerical precision by the diagonal matrix.
If we multiply the modified matrices  $U\times D\times T$, we recover the original, singular square matrix $B$.
%

%%%%%%%%%%%%%%%%%%%%%%%%%%%%%%%%%%%%%%%%%%%%%%%%%%%%%%%%%%%%%%%%%%%%%%
% Stable Green's
%%%%%%%%%%%%%%%%%%%%%%%%%%%%%%%%%%%%%%%%%%%%%%%%%%%%%%%%%%%%%%%%%%%%%%
%
\subsection{Stable calculation of the Green's function}
We now turn to the stable calculation of the equal-time Green's function, which is calculated from the inverse of the matrix product $\mc{B}(\tau)$,
see Eq.~\eqref{eq:greens}, and therefore also sensitively depends on the matrix decomposition discussed in the previous section. 
Despite having circumvented the problem of singularity, the calculation of the Green's functions remains difficult, because the matrices of the decomposition typically remain ill-conditioned, i.e. they retain an extremely broad singular value spectrum resulting in a high condition number.
The problem of calculating the Green's function from ill-conditioned matrices in the finite-temperature algorithm has long been known and was solved by Hirsch and Fye \cite{hirsch_anderson_1988} by using {\it multiple} consecutive matrix decompositions (instead of just one as given in Eq.~\eqref{eq:greens}). These multiple decompositions can be arranged in an enlarged matrix of size $(N_{\text{sites}}\cdot N_{\text{decompositions}}) \times (N_{\text{sites}}\cdot N_{\text{decompositions}})$, such that the determinant of this enlarged matrix remains equal to that of the original one, but  the equal-time Green's function can now be read off as a submatrix of the inverse of the enlarged matrix.
The deeper reason that this approach allows to avoid the ill-conditioned matrix problem above is found in a considerably narrower singular value spectrum of the enlarged matrix. 
A maximum matrix size for the enlarged matrix is reached when each of the slice matrices is used as an input decomposition, i.e. $N_{\text{decompositions}} = N_\tau$. However, it is prohibitively expensive to invert such a large matrix. Fortunately, for our entanglement computations at hand it typically suffices to choose just two or three consecutive decompositions for the calculation of $G_0$ (for a simulation cell with a complete cut) and up to five for $G_1$ (for a simulation cell with a partial cut).

One more technical caveat in the entanglement computation that warrants attention arises when inserting the density matrices $\ketbra{\psi_T}{\psi_T}$ in the calculation of the trace over the reduced density matrix in Eq.~\eqref{eq:replica_gs}.
Upon building a matrix product like $\mc{B} = B(\theta, \theta / 2º)\ketbra{\psi_T}{\psi_t}B(\theta/2, 0)$ we start from the right and multiply slice matrices $\mat{B}(i)$, applying the stabilization procedure using successive matrix decompositions, until we built up the decomposition for the slice matrix group $B(\theta/2, 0)$. 
Up until this point in imaginary time, these matrices are square and invertible but now the insertion of the singular density matrix turns the entire product into a singular matrix. 
If  we now carried on multiplying slice matrices on the left and decomposing the resulting matrices as we had done before, we would expect to find $N_p$ non-zero singular values, corresponding to the particle number of the trial wave function and $N - N_p$ zero singular values.
In practice, however, a decomposition algorithm like SVD or QR will typically find only the $N_p$ non-zero singular values to high precision but the $N - N_p$ remaining singular values (strictly zero in theory) are found to be zero only relative to the actual non-zero singular values.
These inaccuracies will accumulate as we keep multiplying more slice groups and decomposing the resulting matrix products which ultimately results in incorrect Green's matrices. 

This problem can be overcome by a modification of the stack structure, whose implementation we discuss in the following for the case of calculating $G_0$.
In this modified stack structure, we keep track of {\it three} such stacks: one which is built from the bra version of the wave function and includes decompositions obtained from matrices of the form $\bra{\psi}B(\tau, \tau^\prime)$, a second one which is simply the ket version based on $B(\tau, \tau^\prime)\ket{\psi}$ and finally a third one that is used as temporary storage and includes decompositions of the full slice matrix groups $B(\tau, \tau^\prime)$. 
In combination, these three stacks allow to calculate $G_0$ for a given imaginary time slice (at imaginary time $n \cdot \Delta$, see section \ref{sec:Numerics}). 
An example configuration of the three stacks for imaginary time $\tau = n \cdot \Delta$ looks as follows
\begin{figure}[htp]
\includegraphics[width=\columnwidth]{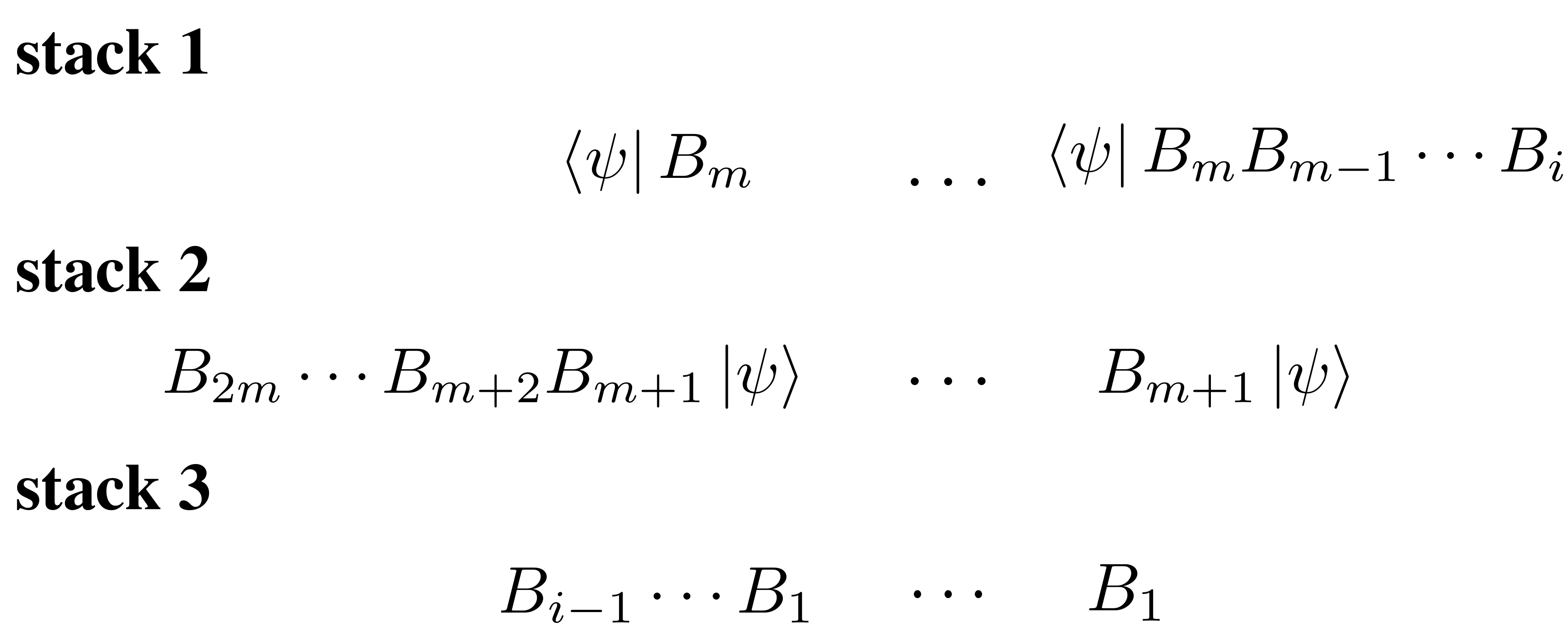}
\end{figure}\\
These three stacks can be used to set up a $3\cdot N_{\text{sites}} \times 3\cdot N_{\text{sites}}$ matrix and  to subsequently calculate the Green's function and the corresponding Monte Carlo weight.
Alternatively, if the condition numbers of the matrices allow it, it is possible to contract two or even all of them for a faster calculation. 
In that case, one has to think about the order of the matrix contractions. 
Naturally, the optimal choice is the one that keeps the condition numbers for the resulting matrices as small as possible, thus ensuring the highest stability of the following operations.
Note that the number of contractions can vary for each recalculation and should be chosen according to the magnitude of the relative error between propagated and recalculated Green's function as defined in Eq.~\eqref{eq:greens_error}.

\begin{figure}
\includegraphics[width=\columnwidth]{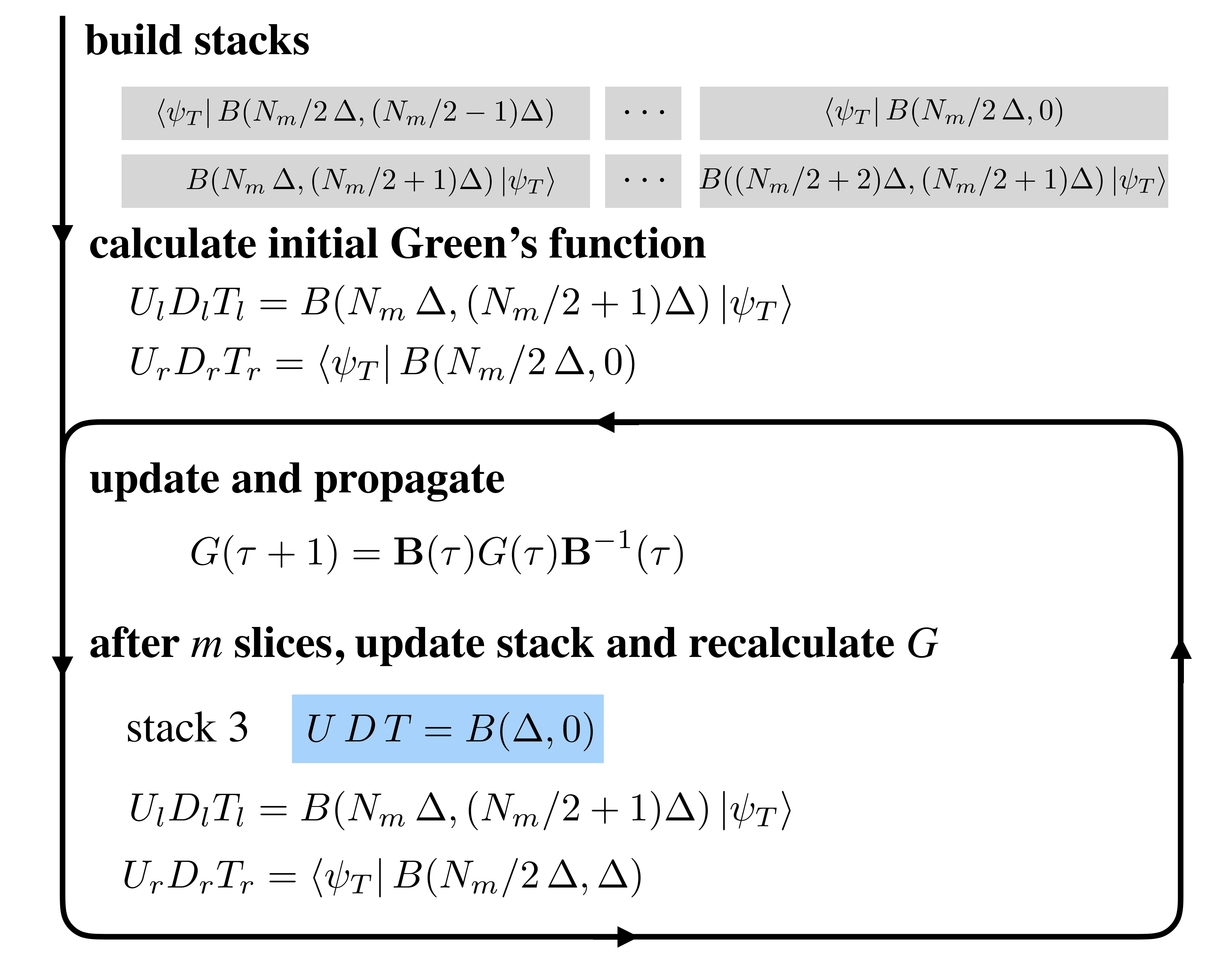}
\caption{(Color online) Flow diagram of the modified DQMC algorithm for the simulation of the partition function $Z_0$ with complete cut in imaginary time. 
Stacks are initialized starting from the bra and ket version of the wave function and subsequently used to calculate an initial Green's function. 
This Green's function is then used to update the auxiliary field time slice by time slice and recalculated after $m$ time steps to retain numerical stability. Intermediate decompositions of slice matrix groups $U, D, T$ without wave function are stored in a third stack, depicted in blue.
		\label{fig:dqmc_replica_scheme}}
\end{figure}

These numerical refinements are incorporated in the modified flow diagram of our DQMC algorithm in Fig.~\ref{fig:dqmc_replica_scheme}.
As for the calculation of $G_1$, we have to keep track of five individual stacks but the rest of the algorithm works exactly as just described for $G_0$.

%%%%%%%%%%%%%%%%%%%%%%%%%%%%%%%%%%%%%%%%%%%%%%%%%%%%%%%%%%%%%%%%%%%%%%
% Choosing the right Hubbard-Stratonovich transformation
%%%%%%%%%%%%%%%%%%%%%%%%%%%%%%%%%%%%%%%%%%%%%%%%%%%%%%%%%%%%%%%%%%%%%%

\subsection{Choice of Hubbard-Stratonovich transformation}
\label{sec:HS}

\begin{figure*}[htp!]
  \subfigure{\includegraphics[width=\columnwidth]{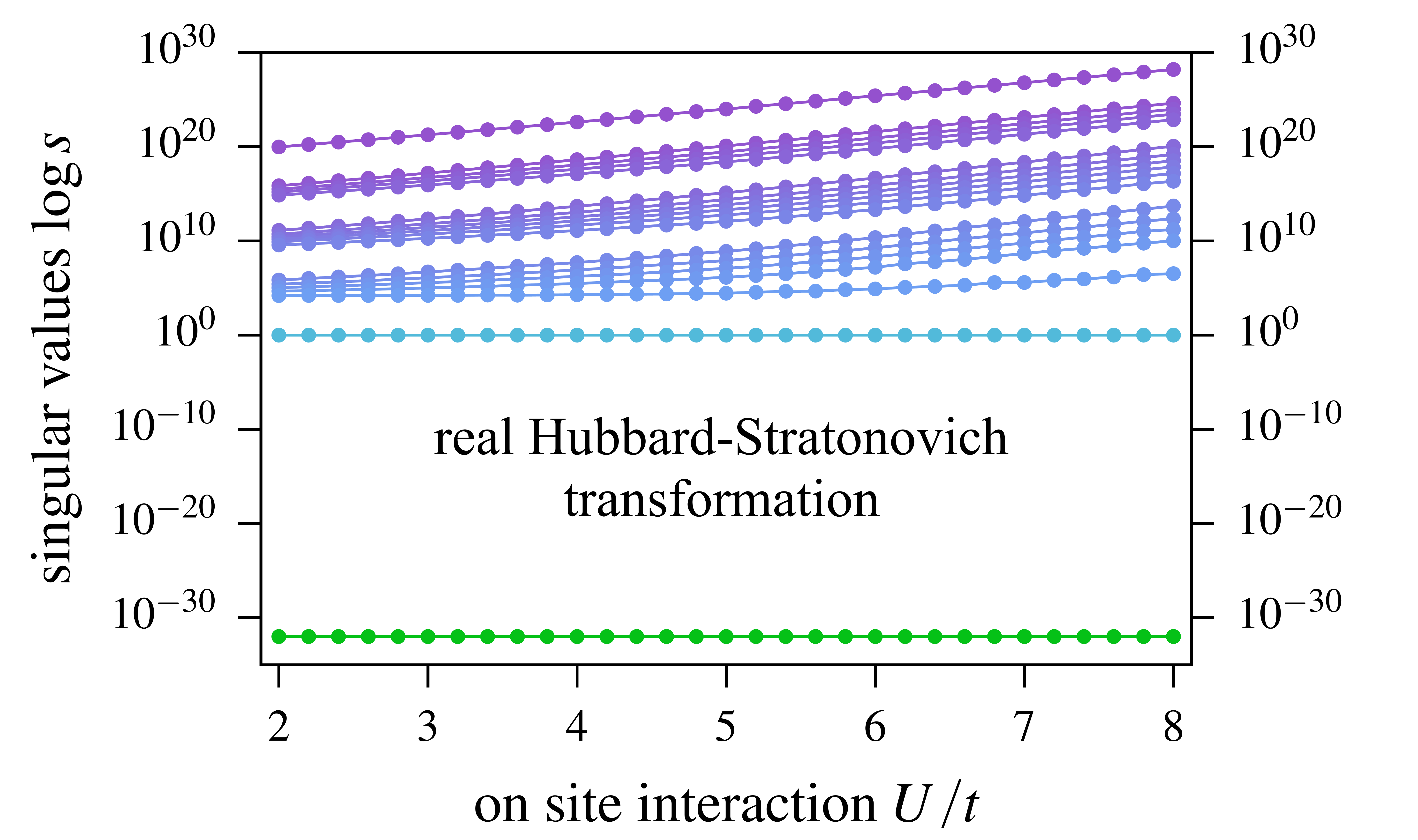}}\quad
  \subfigure{\includegraphics[width=\columnwidth]{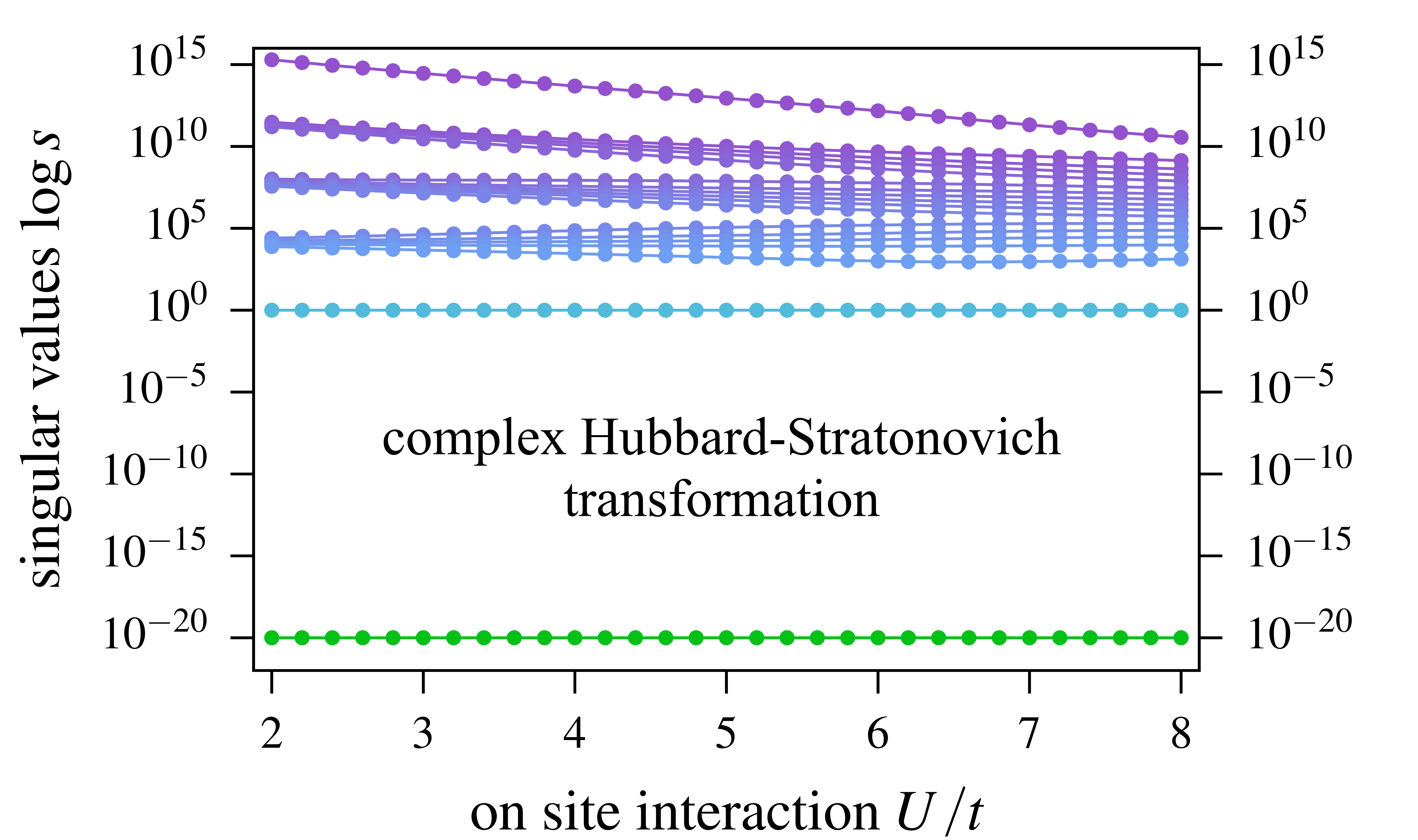}}
  \caption{(Color online) The singular value spectrum of the matrix $B(\Theta / 2, 0)\ket{\psi_T}$ i.e. the projected wave function, on a logarithmic scale for a real (left) and complex (right) Hubbard-Stratonovich transformation.\label{fig:comparison_hs}
  		  The projection time is $\theta = 10$.
		  }
		
\end{figure*}
At the heart of the DQMC approach is the decoupling of quartic terms in the Hamiltonian using a Hubbard-Stratonovich transformation. 
In general, there is a multitude of possible alternatives to perform this transformation. Depending on the physics of interest, common choices include a decoupling in the spin or charge channels which leads to either a real or complex Hubbard-Stratonovich transformation that preserves or breaks SU(2) symmetry, respectively. From an algorithmic point of view, the choice of Hubbard-Stratonovich transformation can greatly affect
the numerical stability and convergence of the DQMC approach.

To illustrate the influence of the choice of Hubbard-Stratonovich transformation on the algorithmic performance, let us consider the example of a density-density interaction of the general form
\begin{equation}
	\hat{U}_{\alpha\beta} = U_{\alpha\beta}\, n_\alpha n_\beta \,,
	\label{eq:interaction}
\end{equation}
where the indices $\alpha$ and $\beta$ may represent adjacent sites for a nearest-neighbor interaction or different spin species for a given site.
In either case, we are looking to identify a quadratic operator $A$ such that we can apply the Hubbard-Stratonovich transformation 
\begin{equation}
	e^{\frac{1}{2}A^2} = \sqrt{2\pi}\int\limits_{-\infty}^{\infty}\textrm{d}s\; e^{-\frac{1}{2}s^2 - s\,A} \,,
	\label{eq:general_hs}
\end{equation}
where $s$ is the auxiliary field introduced in the decoupling scheme.
To do so, we will rewrite the interaction term \eqref{eq:interaction} in an appropriate form. There are two possible choices
\begin{align}
	n_\alpha n_\beta &= -\dfrac{1}{2}(n_\alpha - n_\beta)^2 + \dfrac{1}{2}(n_\alpha^2 + n_\beta^2) \,, \label{eq:squaredform1} \\
	n_\alpha n_\beta &= \dfrac{1}{2}(n_\alpha + n_\beta)^2 - \dfrac{1}{2}(n_\alpha^2 + n_\beta^2) \,, \label{eq:squaredform2}
\end{align}
with the quadratic operator corresponding to the first term, i.e. $A=(n_\alpha - n_\beta)^2$ or $A=(n_\alpha + n_\beta)^2$, respectively. For the scenario of a spin-spin onsite-interaction the first choice of $A$ in Eq.~\eqref{eq:squaredform1} corresponds to the squared magnetization and thus a decoupling in the spin channel. For the second choice \eqref{eq:squaredform2} the operator $A$ corresponds to the squared total charge and thus corresponds to a decoupling in the charge channel.

Note that in order to perform the Hubbard-Stratonovich transformation \eqref{eq:general_hs}, the auxiliary field $s$ does not necessarily have to be continuous, as probably suggested by the integral form of Eq.~\eqref{eq:general_hs}. For some cases, it suffices to work with a discrete  field that is constrained to, e.g., take values $\pm 1$ such as the case of the spinful Hubbard model, with most applications typically needing no more than a four-valued discrete field \cite{assaad_charge_1997-1}.

Applying the Hubbard-Stratonovich transformation \eqref{eq:general_hs} after bringing the operators into squared form (\ref{eq:squaredform1}, \ref{eq:squaredform2}) gives the following expressions
\begin{equation}
e^{-\Delta\tau U n_\alpha n_\beta} = \dfrac{1}{2}\sum\limits_{s = \pm 1}\prod\limits_{a = \alpha,\beta} e^{-\left(s\lambda  + \frac{U\Delta\tau}{2}\right)n_a} \label{eq:hs_repulsive}
\end{equation}
and
\begin{equation}
e^{-\Delta\tau U n_\alpha n_\beta} = \dfrac{1}{2}\sum\limits_{s = \pm 1}\prod\limits_{a = \alpha,\beta} e^{-\left(s\lambda  + \frac{U\Delta\tau}{2}\right)\left(n_a - \frac{1}{2}\right)} \,, \label{eq:hs_attractive}
\end{equation}
respectively.
Here the constant $\lambda$ is given by
\begin{equation}
	\cosh{\left(\lambda\right)} = e^{\frac{1}{2}U\Delta\tau}
	\label{eq:lambda_repulsive}
\end{equation}
and 
\begin{equation}
	\cosh{\left(\lambda\right)} = e^{-\frac{1}{2}U\Delta\tau} \,,
	\label{eq:lambda_attractive}
\end{equation}
respectively. Note that Eqs.~(\ref{eq:lambda_repulsive},\ref{eq:lambda_attractive}) imply that depending on the sign of the interaction $U$, the constant $\lambda$ can either be real or complex valued. 

At first look, one might think that the real-valued Hubbard-Stratonovich transformation leads to lower computational cost. However, in computations of the entanglement entropy  it is the complex-valued Hubbard-Stratonovich transformation that should be used preferably. The reason for this preference is that the type of Hubbard-Stratonovich transformation also greatly affects the singular value spectrum of the matrix decompositions discussed in the previous subsection.
To illustrate this point, we show in Fig.~\ref{fig:comparison_hs} a comparison of the singular value spectrum of the matrix $B(\Theta / 2, 0)\ket{\psi_T}$, i.e. the projected wave function, obtained for real and complex Hubbard-Stratonovich transformations. The data is calculated for a bilayer Hubbard model on a square lattice of size $4\times 4\times 2$ at half filling and equal hopping within and between the layers, i.e. $t = t^\prime$, but varying onsite interaction $U$, see also Sec.~\ref{sec:Bilayer} for a more detailed discussion of this model.
In the complex case, the range of the singular values is seen to decrease as the interaction increases, hence reducing the condition number and stabilizing the algorithm.
In the real case, the behavior is found to be exactly opposite with the condition number becoming worse as the  interaction strength increases.

As a consequence, we always use complex-valued Hubbard-Stratonovich transformation in the computation of \Renyi entanglement entropies.

%%%%%%%%%%%%%%%%%%%%%%%%%%%%%%%%%%%%%%%%%%%%%%%%%%%%%%%%%%%%%%%%%%%%%%
% Convergence
%%%%%%%%%%%%%%%%%%%%%%%%%%%%%%%%%%%%%%%%%%%%%%%%%%%%%%%%%%%%%%%%%%%%%%

\subsection{Convergence}
\label{sec:Convergence}

%%%%%%%%%%%%%%%%%%%%%%%%%%%%%%%%%%%%%%%%%%%%%%%%%%%%%%%%%%%%%%%%%%%%%%
% Artificial chemical potential
%%%%%%%%%%%%%%%%%%%%%%%%%%%%%%%%%%%%%%%%%%%%%%%%%%%%%%%%%%%%%%%%%%%%%%

\begin{figure}
\includegraphics[width=\columnwidth]{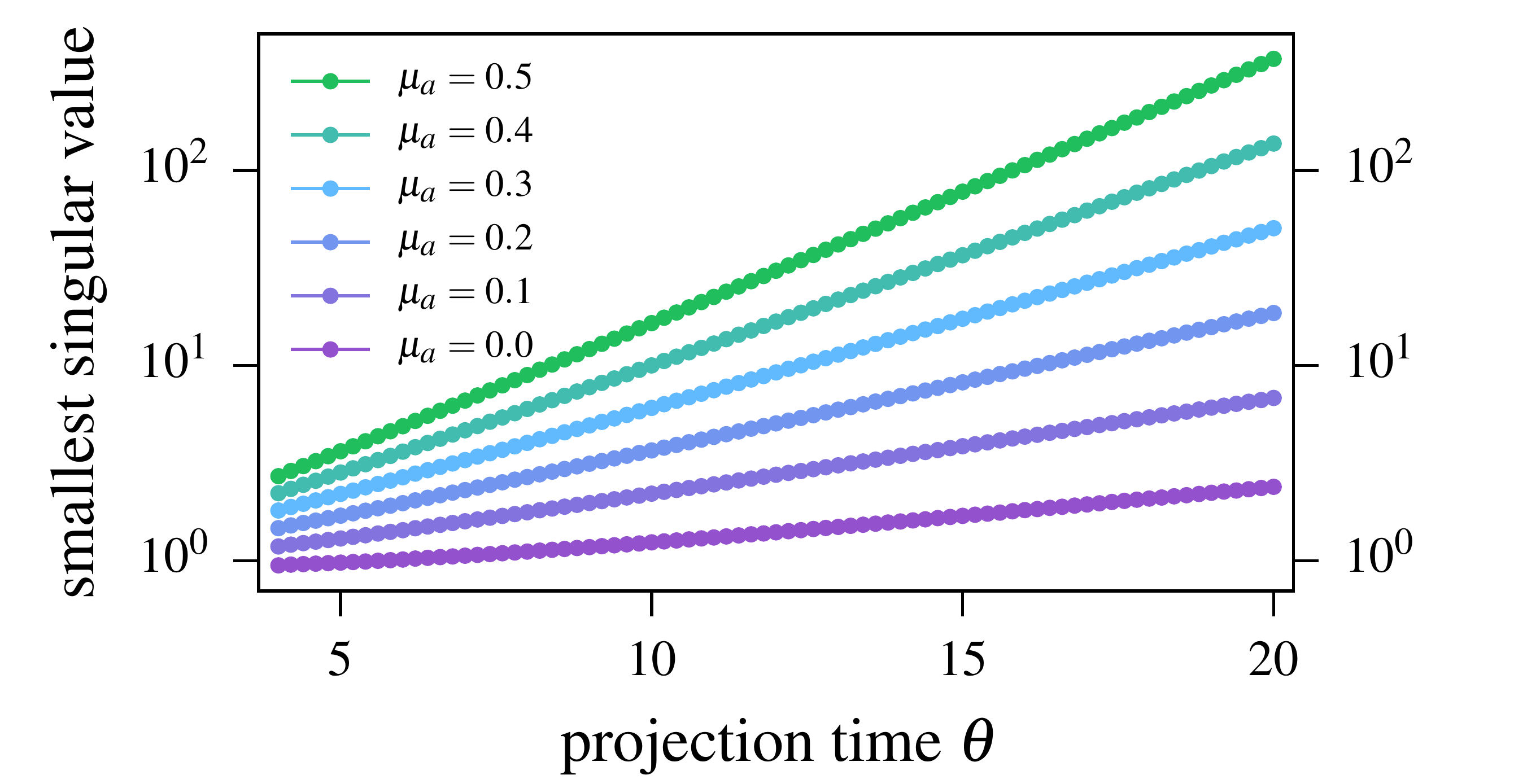}
\caption{(Color online) Evolution of the smallest singular value with projection time $\Theta$ 
		for various values for the artificial chemical potential $\mu_a$.
		\label{fig:convergence_mu_singular_value_gap}}
\end{figure}

\begin{figure}
\includegraphics[width=\columnwidth]{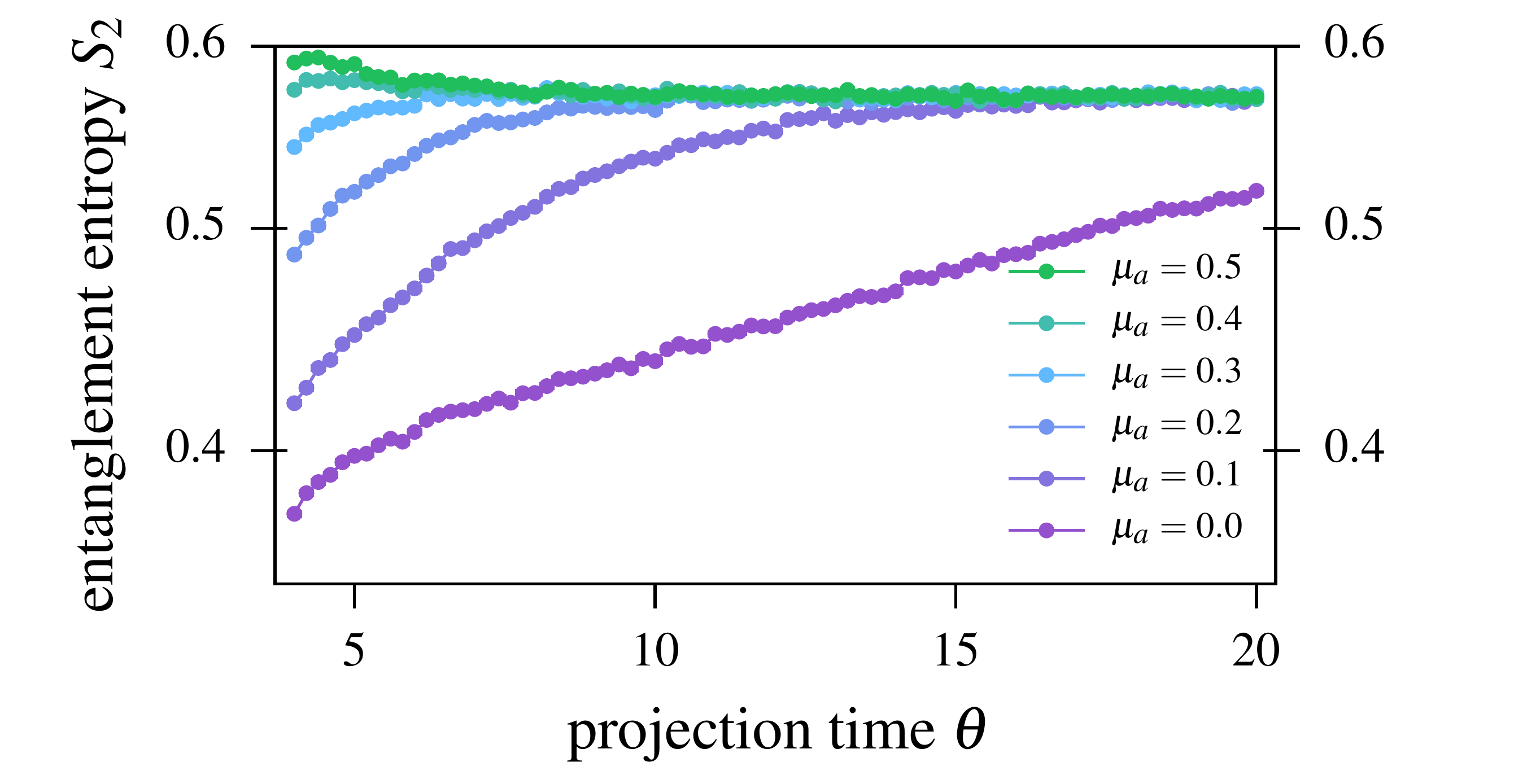}
\caption{(Color online) Convergence of the \Renyi entanglement entropy $S_2$ in the projective DQMC algorithm versus projection time for varying 
		values of the artificial chemical potential $\mu_a$. The \Renyi entropy is calculated for a half-filled Hubbard model
		on the square lattice for $U/t=4$ with an equal-size bipartition of the lattice.		
		\label{fig:convergence_mu}}
\end{figure}

One key distinction of our zero-temperature entanglement calculations in the projective DQMC approach outlined in this manuscript and the closely related finite-temperature algorithm is in the type of simulated ensemble. 
While the finite-temperature algorithm is sampling states from a grand canonical ensemble, the projective scheme samples a canonical ensemble of {\it fixed} particle number (encoded in the trial wave function). 
This seemingly small modification of the ensemble is found to have some rather noticeable impact on the convergence properties of the projective algorithm. To ensure convergence, we resort to a technical trick by introducing an artificial chemical potential in the Hamiltonian
\begin{equation}
	H^\prime = H + \mu_a \sum\limits_i n_i \,.
	\label{eq:artificialpotential}
\end{equation}
Note that the inclusion of this artificial chemical potential $\mu_a$ is not altering the physics of the original problem, since we keep working in a canonical
ensemble with a fixed particle number.
To understand why it is nevertheless beneficial to include this term, let us first note that, on a technical level, such a term is represented by a diagonal matrix that has the ability to significantly shift the singular value spectrum.
This is displayed in Fig.~\ref{fig:convergence_mu_singular_value_gap}, which plots the smallest singular value of the unmodified matrix $B(\Theta / 2, 0)\ket{\psi}$. 
Clearly visible is the increase of the magnitude of the smallest singular value with increasing projection time $\Theta$ and increasing chemical potential $ \mu_a$.
If we now consider the convergence of the entanglement entropy upon inclusion of this artificial chemical potential term, we find that already a small additional potential dramatically improves the convergence of the entanglement entropy as a function of projection time as shown in Fig.~\ref{fig:convergence_mu}. Further increasing the artificial chemical potential we find that the convergence eventually saturates, in the case at hand for about $ \mu_a \approx 0.4$. In general, the optimal value of $ \mu_a$ depends rather sensitively on the parameters of the Hamiltonian and needs to be chosen with great care.

\begin{figure}[b]
\includegraphics[width=\columnwidth]{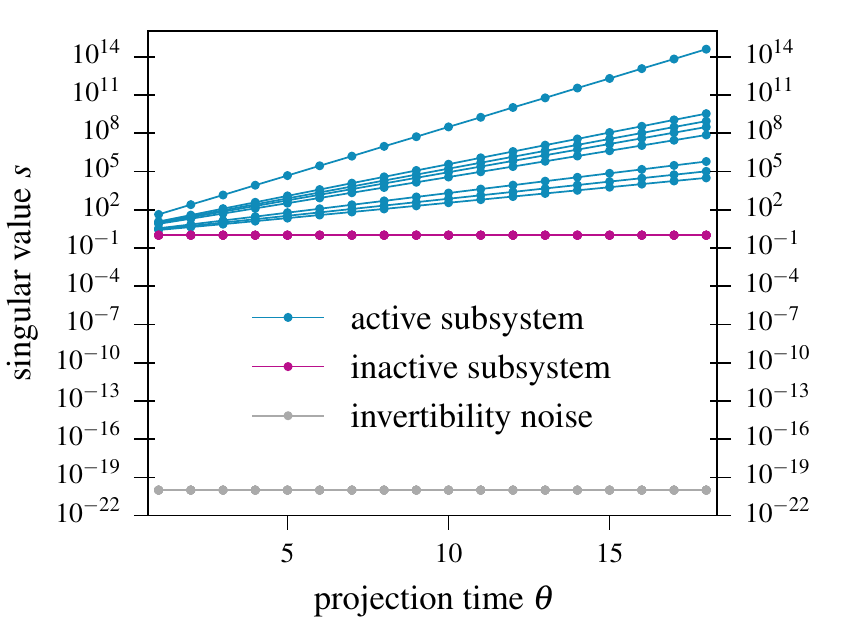}
\caption{(Color online) Magnitude of the singular values of the Green's function in the replicated system for a Hubbard model on a square lattice at half filling, an on site interaction of $U/t=4$ and $\mua = 0.5$.
		\label{fig:singular_values_theta}}
\end{figure}

Closer inspection of the convergence properties of our entanglement DQMC algorithm reveals that the added chemical potential \eqref{eq:artificialpotential} only affects the calculation of the Green's function $G_1$ associated with the partition sum $Z_1$, i.e.  the partition sum for the partially cut system, see Fig.~\ref{fig:ensemble_switching_paths}. 
This observation already hints at the origin of the improved convergence. For the partition sum $Z_1$ all slice matrices are obtained
from the ``unfolded" Hamiltonian \eqref{eq:img_time_ham}, where we fold out subsystem $B$ as illustrated in Fig.~\ref{fig:ensemble_switching_unfolded}. The slice matrices include blocks of identity and zero matrices to accommodate the currently ``inactive" subsystem $B$ or $B^\prime$, i.e. the one which is deactivated by the $\Theta$-function in the Hamiltonian \eqref{eq:img_time_ham}.
As demonstrated in Fig.~\ref{fig:singular_values_theta} this greatly affects the singular value spectrum by introducing $N_B$ singular values of unit value (indicated by the magenta data set) representing the said subsystem.
Comparing Figs.~\ref{fig:convergence_mu_singular_value_gap} and \ref{fig:singular_values_theta}, we observe that once the difference between the smallest singular value of the the active subsystem becomes separated from the unit singular values corresponding to the inactive subsystem, the simulation has a chance to converge.
Thus, the requirement is that this gap $\Delta_s$ in the singular value spectrum has to be sufficiently large. In our studies we empirically find that $\Delta_s \approx 10^2$ turns out to be a good choice for all calculations.

%%%%%%%%%%%%%%%%%%%%%%%%%%%%%%%%%%%%%%%%%%%%%%%%%%%%%%%%%%%%%%%%%%%%%%
% Phase transition in the bilayer Hubbard model
%%%%%%%%%%%%%%%%%%%%%%%%%%%%%%%%%%%%%%%%%%%%%%%%%%%%%%%%%%%%%%%%%%%%%%

\section{Application to the bilayer Hubbard model}
\label{sec:Bilayer}

We round off the technical discussion of this manuscript with one example illustrating the application of entanglement computations to identify
quantum phases of interacting many-fermion systems. We resort to the bilayer Hubbard model -- a paradigmatic model that allows to study the transition between a Mott insulator and a band insulator. Whether these two insulating states are fundamentally distinct or can
in general be adiabatically connected into one another \cite{Anfuso2007}, has been a question of debate \cite{Essler2002,Essler2005,Monien2006,Konik2006,Berthod2006,Garg2006,Stanescu2007,Rosch2007,Okamoto2007,Scalettar2007,Dagotto2007}. This discussion mostly preceded the days of the topological insulator \cite{Hasan2010,Qi2011} -- a second type of band insulator that can be clearly distinguished from the conventional ``trivial" band insulator by certain topological invariants \cite{Kane2005a,Kane2005b}, while its fundamental distinction to correlated Mott insulators (possibly exhibiting topological order as well) is an open question of much current interest \cite{Senthil2014}.

\begin{figure}[t]
\includegraphics[width=\columnwidth]{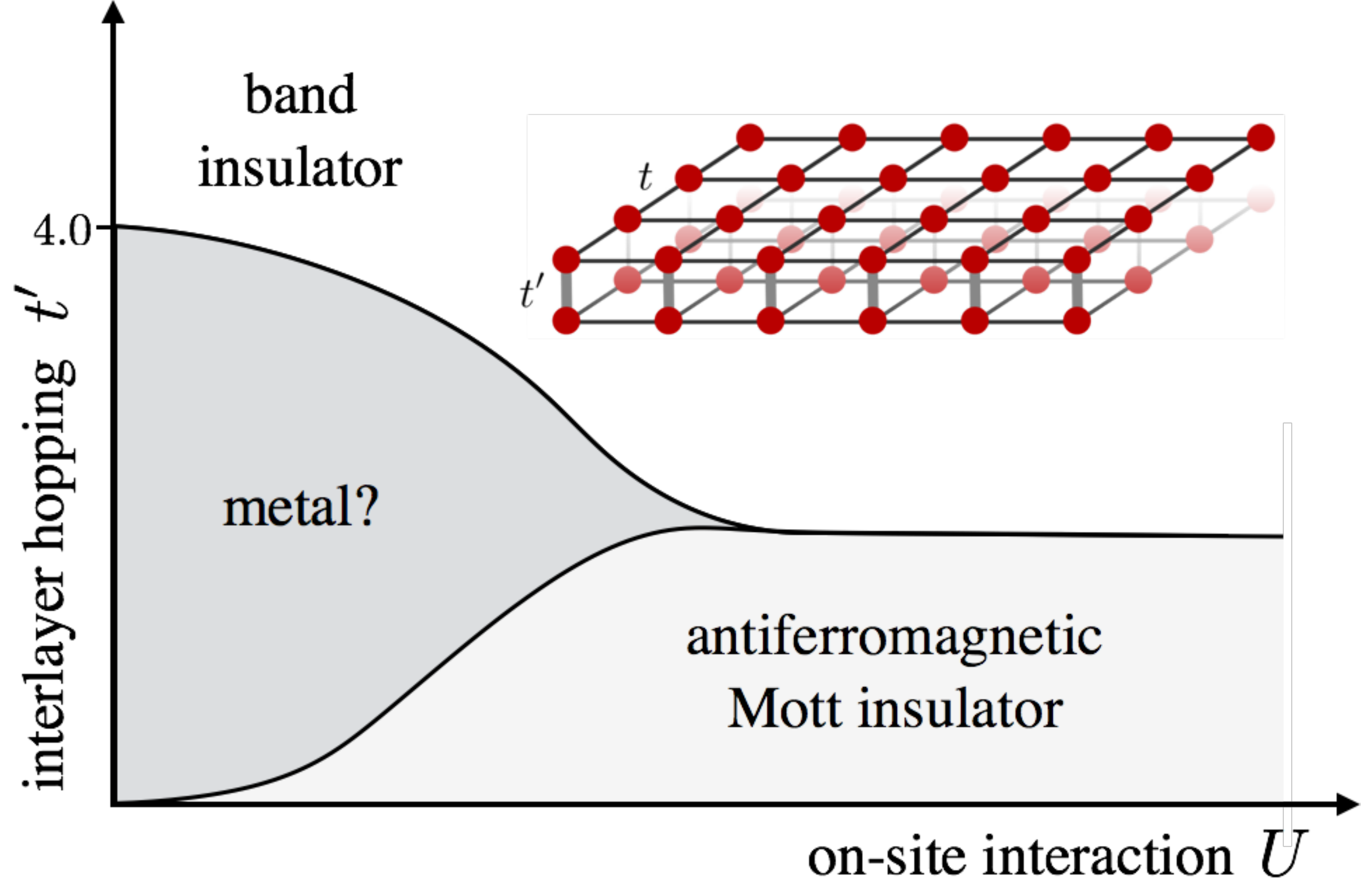}
	\caption{(Color online) Schematic phase diagram for the bilayer Hubbard model at half filling, 
		adapted from dynamical mean field theory (DMFT) \cite{Okamoto2007},
		DQMC simulations \cite{Scalettar2014}, and functional renormalization group (FRG) calculations \cite{Golor2014}.
		It comprises two insulating phases, a band insulator for sufficiently strong interlayer hopping $t^\prime \gtrsim 4$,
		an antiferromagnetic insulator for weak interlayer hopping strength and large on-site Coulomb repulsion $U$.
		There possibly is an extended metallic phase in the regime of small Coulomb repulsion and interlayer hopping
		as indicated in the phase diagram, with DMFT and DQMC simulations in favor of it \cite{Okamoto2007,Scalettar2014}, 
		while FRG calculations point to an absence of a metallic phase for any finite Coulomb repulsion \cite{Golor2014}.
		The inset depicts of the bilayer lattice. The intralayer hoppings are indicated by $t$, 
		while $t^\prime$ denotes the interlayer hoppings.
	\label{fig:phase_diagram_cartoon}}
\end{figure}

Here we want to apply an entanglement perspective on the elementary phase diagram of the bilayer Hubbard model 
\begin{equation}
H = -t\sum\limits_{\langle i, j\rangle, \sigma} c^{\dagger}_{i, \sigma} c_{j, \sigma} 
-t^\prime\sum\limits_{\langle i, j\rangle^\prime, \sigma} c^\dagger_{i, \sigma} c_{j, \sigma} 
+ U\sum\limits_{i} n_{i, \uparrow} n_{j, \downarrow} \,,
\end{equation}
which at its core describes the competition between conventional (free-fermion) band structures arising from hopping within and between the 
layer of a double-layer square lattice (parametrized by hopping amplitudes $t$ and $t^\prime$, respectively) and Mott physics arising from an on-site Coulomb repulsion $U$.
A schematic phase diagram for this model in terms of the on-site interaction $U$ and the interlayer hopping $t^\prime$ is given in Fig.~\ref{fig:phase_diagram_cartoon}. For sufficiently large interlayer hopping strength $t^\prime$ and any value of $U$ the system is a 
featureless band insulator, while for sufficiently large $U$ and small interlayer hopping $t^\prime$ the system forms a Mott insulator
with antiferromagnetic order. The phase transition from Mott to band insulator has been studied using a variety of methods in the past~\cite{weihong_various_1997, wang_high_precision_2006, kancharla_band_2007, scalettar_magnetic_2008, hamer_restoration_2012, helmes_entanglement_2014, rueger_phase_diagram_2014, Scalettar2014,Golor2014}.
In the absence of a Hubbard interaction $U$ and small interlayer hopping we have a metal.
Whether this metallic state survives for small Hubbard $U$ (and small interlayer hopping $t^\prime$ ) or immediately gives way to Mott physics
is still under debate, with dynamical mean field theory (DMFT) \cite{Okamoto2007} and DQMC simulations \cite{Scalettar2014} pointing to an extended metallic phase, while a functional renormalization group (FRG) analysis \cite{Golor2014} finds an immediate breakdown of the metallic phase upon inclusion of the Hubbard term. 

Our DQMC approach is ideally suited to compute \Renyi entanglement entropies over a wide range of parameters, in particular going deep into the regime of strong coupling $U > t,t^\prime$. To do so, we consider a bipartition of the system into a strip of width $L/4 \times L \times 2$ and its complement. For this type of cut, which includes both the upper and lower layer, we expect that the entanglement signature of the band insulator is rather trivial -- with the formation of singlet dimers on the rungs between the layers any cut between the rungs will effectively see no significant entanglement contribution, similar to the entanglement signature of an unentangled product state. Deep in the Mott regime we expect to observe the prevalent entanglement signature of any entangled quantum many-body state -- a finite entanglement entropy that is subject to the famous boundary-law scaling \cite{Eisert2010}. For any metallic state we expect to observe an even stronger entanglement signature in the form of a logairthmic violation of the aforementioned boundary law \cite{Wolf2006,Klich2006}.

\begin{figure}[t]
  \includegraphics[width=\columnwidth]{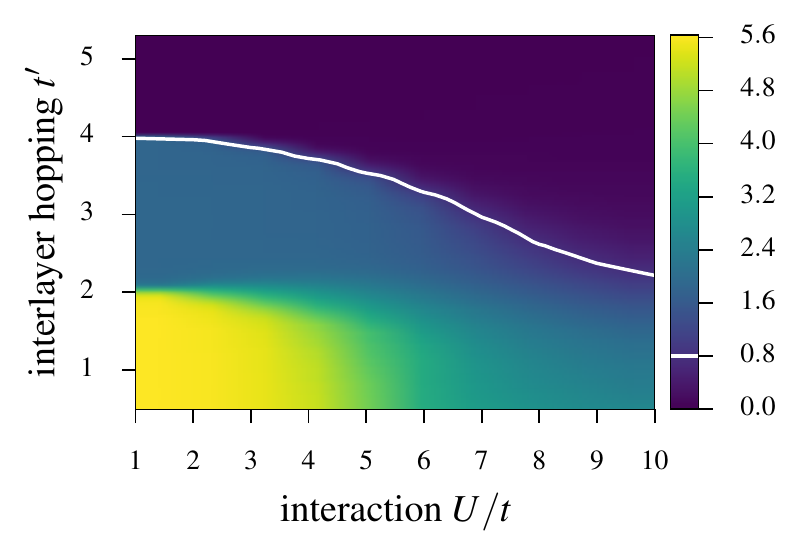}
  \caption{(Color online) Color-coded \Renyi entanglement entropy $S_2$ for the bilayer Hubbard model calculated for a bipartition 
  		of a $4 \times 4 \times 2$ system (with the system divided into a strip of extent $1 \times 4 \times 2$ and
		its complement). The different entanglement regimes clearly reveal the general characteristics
		of the underlying phase diagram. For dominant interlayer coupling $t^\prime \gtrsim 4$ the vanishingly small
		entanglement points to the featureless band insulator with singlet formation on the interlayer rungs. 
		For moderate interlayer coupling, the finite entanglement reveals the Mott insulator. 
		For small interlayer coupling and small onsite interaction, a third entanglement regime can be distinguished
		(indicated by the yellow region), which possibly points to an additional metallic phase.  		
		The phase boundary separating the band insulator is indicated by an equipotential line of $S_2 = 0.8$. 
		The relevant color values corresponding to these values of the entanglement entropy are also marked on the colorbar by white lines. 
  		\label{fig:phase_diagram_dqmc}}
\end{figure}

\begin{figure}[t!]
\includegraphics[width=\columnwidth]{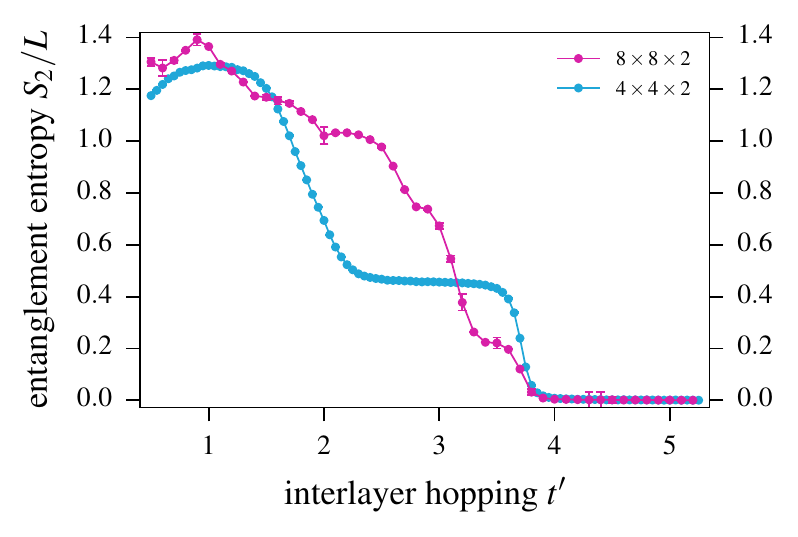}
\caption{(Color online) \Renyi entanglement entropy $S_2$ for the bilayer Hubbard model along a vertical cut through the
		phase diagram of Fig.~\ref{fig:phase_diagram_dqmc} at $U/t=4$. 
		Data for different system sizes is shown. 
		\label{fig:s2_bilayer}}
\end{figure}

As a first step we have scanned the {\it absolute} value of the \Renyi entropy for $t^\prime \in \left[0.5, 5.0\right]$
as illustrated in Fig.~\ref{fig:phase_diagram_dqmc} for a system of $4 \times 4 \times 2$ sites. Clearly, several distinct regimes
of almost constant amount of entanglement can be readily distinguished thereby revealing some of the core features of the phase diagram.
For instance, one can clearly identify a regime of almost vanishing entanglement for large interlayer coupling $t^\prime$ -- this is the band insulator as argued above. The finite entanglement clearly present in the lower part of the phase diagram points to at least one separate regime, possibly two when considering the step-like enhancement of the \Renyi entropy for  $t^\prime/t < 2$ and $U/t \lesssim 4$ indicated by the yellow color. The \Renyi entropy along a vertical cut through the data of Fig.~\ref{fig:phase_diagram_dqmc} is shown in Fig.~\ref{fig:s2_bilayer} for $U/t=4$.

\begin{figure}[t!]
\includegraphics[width=\columnwidth]{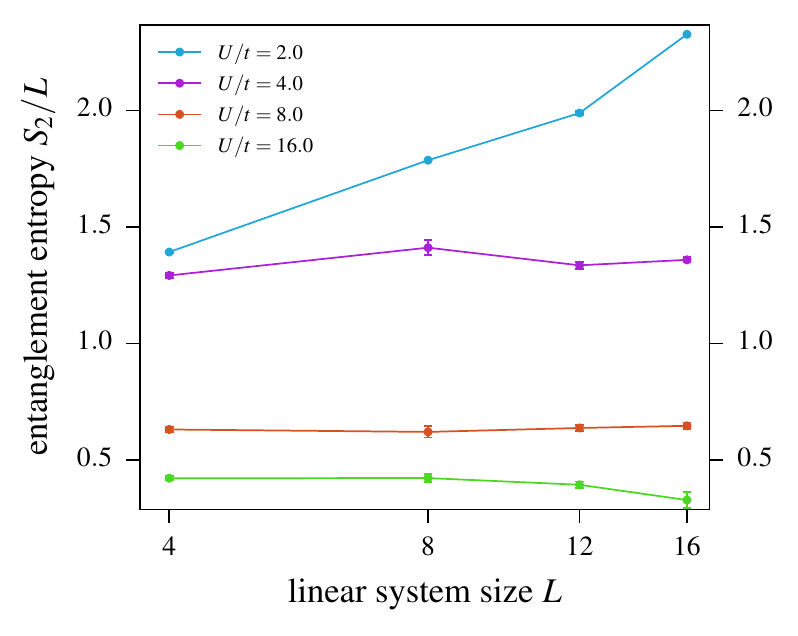}
\caption{(Color online) Finite-size scaling of the entanglement entropy $S_2$ renormalized by the linear system size $L$ 
		for the bilayer Hubbard model at different values of the on-site interaction $U/t$ and fixed interlayer coupling $t^\prime/t=1$.
		Note the logarithmic scale of the abscissa.
		\label{fig:s2_scaling}}
\end{figure}

To further substantiate the nature of the phase(s) in the lower half of the phase diagram we consider the scaling of the \Renyi entropy with system size. In Fig.~\ref{fig:s2_scaling} we show results for the \Renyi entropy for different values of the onsite Coulomb repulsion $U/t=2,4,8$ and $16$  and fixed interlayer hopping $t^\prime/t =1$.
With the entropy  renormalized by the linear system size $L$, the boundary law scaling expected for the Mott insulator corresponds to a flat line. This is precisely what we find for moderate to large Hubbard interaction $U/t = 4, 8, 16$. In contrast, for small Hubbard interaction $U/t=2$ the clearly noticeable slope of the data points suggests an additional logarithmic contribution to the entanglement entropy. Precisely such an $S \propto L \log L$ scaling is expected for a two-dimensional metal with a Fermi surface \cite{Wolf2006,Klich2006}. Note, however, that we are looking at rather small system sizes when inferring this logarithmic contribution and it would be highly desirable to go to substantially larger system sizes to exclude possible finite-size effects. Unfortunately, such a substantial increase in system sizes does not seem feasible with the current algorithms and state-of-the-art computational resources. Nevertheless, with the data at hand the entanglement point of view is certainly adding support to a scenario, in which the metallic state survives the inclusion of a moderately small Hubbard interaction. 

%

%%%%%%%%%%%%%%%%%%%%%%%%%%%%%%%%%%%%%%%%%%%%%%%%%%%%%%%%%%%%%%%%%%%%%%
% Conclusions
%%%%%%%%%%%%%%%%%%%%%%%%%%%%%%%%%%%%%%%%%%%%%%%%%%%%%%%%%%%%%%%%%%%%%%

\section{Summary}
\label{sec:Conclusions}

To summarize, we have laid out a number of algorithmic advances to overcome the numerical instabilities of computing entanglement
entropies for interacting many-fermion systems within the DQMC framework. The basic algorithm \cite{Broecker2014} is an adaptation 
of the replica scheme to the DQMC approach, which on a technical level converts the finite-temperature algorithm (in the grand-canonical ensemble) to a ground-state projective algorithm (in the canonical ensemble). 
Our numerical improvements of this approach, reported on here,
include (i) a numerical stabilization of the matrix inversion of seemingly singular matrix products via a refined matrix decomposition,
(ii) a numerical stabilization of the calculation of the Green's function using this refined matrix decomposition, 
(iii) an optimal choice of Hubbard-Stratonovich transformation, which typically turns out to be a complex-valued one, and 
(iv) the inclusion of an artificial chemical potential to significantly improve the convergence of the algorithm.
With these algorithmic enhancements implemented, we find that one can reach similar system sizes as for the calculation of more conventional observables in the original projective DQMC scheme and showcased the algorithm by applying it to the bilayer Hubbard model.

Form a physics point of view, probably the most interesting application of the computational scheme at hand is to showcase that 
one can positively identify the formation of macroscopic entanglement in an interacting many-fermion system exhibiting a phase with intrinsic topological order such as the recently proposed models of Dirac fermions coupled to a Z$_2$ gauge field \cite{AssaadGrover2016,Vishwanath2016}. This is ongoing work.

%%%%%%%%%%%%%%%%%%%%%%%%%%%%%%%%%%%%%%%%%%%%%%%%%%%%%%%%%%%%%%%%%%%%%%
% Acknowledgments
%%%%%%%%%%%%%%%%%%%%%%%%%%%%%%%%%%%%%%%%%%%%%%%%%%%%%%%%%%%%%%%%%%%%%%

\acknowledgments
We thank F. Assaad for insightful discussions on the technical aspects of the DQMC approach and S. Wessel for an exchange on the 
conflicting numerical results for the phase diagram of the bilayer Hubbard model. 
P.B. acknowledges partial support from the Deutsche Telekom Stiftung and the Bonn-Cologne Graduate School of Physics and Astronomy (BCGS). We thank the DFG for partial support within the CRC network TR 183 (project B01). The numerical simulations were performed on the CHEOPS cluster at RRZK Cologne and the JURECA cluster at the Forschungszentrum Juelich.

%%%%%%%%%%%%%%%%%%%%%%%%%%%%%%%%%%%%%%%%%%%%%%%%%%%%%%%%%%%%%%%%%%%%%%
% Bibliography
%%%%%%%%%%%%%%%%%%%%%%%%%%%%%%%%%%%%%%%%%%%%%%%%%%%%%%%%%%%%%%%%%%%%%%

%\clearpage
\bibliography{stabilization.bib}

\begin{thebibliography}{74}
\bibitem [{Foo()}]{Footnote:Degeneracy}%
  \BibitemOpen
  \bibinfo {note} {{Note that special care has to be taken when the eigenvalues
  are degenerate \cite{assaad_charge_1997-1}. In that case, one can for example
  randomize the hopping amplitudes a little bit to lift the
  degeneracy}}\BibitemShut {NoStop}%
\end{thebibliography}
\bibliographystyle{bibstyle.bst}

\end{document}